\documentclass[preprint]{revtex4-1}

\usepackage{amssymb}
\usepackage{amsmath,mathtools}
\usepackage{mathrsfs}
\usepackage{graphicx}
\usepackage{epsfig,latexsym}
\usepackage{url}
\usepackage{slashed}
\usepackage{geometry}
\usepackage{simplewick}
\usepackage{caption}
\usepackage{subcaption}
\usepackage{nicefrac}
\usepackage[colorlinks=true,linktocpage=true,linkcolor=blue,citecolor=blue]{hyperref}
\usepackage[usenames,dvipsnames]{color}
\usepackage{lineno}
\modulolinenumbers[5]
\usepackage{xparse}
\usepackage{xargs}
\usepackage[mathscr]{euscript}
\usepackage{appendix}

\def\bs{\boldsymbol} 

\def\bdel{\bs\partial}

\newcommand{\eqn}[1]{Eq.~\eqref{#1}}
\newcommand{\fign}[1]{Fig.~\ref{#1}}
\long\def\comment#1{ }

\newcommand{\nn}{\nonumber\\ }
\newcommand{\onehalf}{{\nicefrac{1}{2}}}

\def\be{\begin{eqnarray*}}
\def\ee{\end{eqnarray*}}
\def\beq{\begin{eqnarray}}
\def\eeq{\end{eqnarray}}
\newcommand{\bea}{\beq \begin{aligned}}
\newcommand{\eea}{\end{aligned}\eeq}

\def\k{{\boldsymbol k}}

\def\p{{\boldsymbol p}}

\def\r{{\boldsymbol r}}

\def\x{{\boldsymbol x}}
\def\y{{\boldsymbol y}}

\def\u{{\boldsymbol u}}
\def\0{{\boldsymbol 0}}

\def\k{{\boldsymbol k}}

\def\n{{\boldsymbol n}}
\def\x{{\boldsymbol x}}
\def\y{{\boldsymbol y}}
\def\p{{\boldsymbol p}}

\def\r{{\boldsymbol r}}

\def\u{{\boldsymbol u}}

\def\Q{{\boldsymbol Q}}

\def\beps{{\boldsymbol \epsilon}}

\def\rme{{\rm e}}
\def\rmd{{\rm d}}
\def\dd{\text{d}}
\newcommand{\rmtr}{{\rm tr}}
\newcommand{\rmTr}{{\rm Tr}}
\def\rmR{{\rm Re}}
\def\rmI{{\rm Im}}
\def\abar{{\rm \bar\alpha}}

\def\Tdc{\Theta_{\rm dc}}
\def\Tdc{\Theta_0}

\def\tmat{ \text{\bf t}}
\def\tmat{ \textbf{t}}
\def\Tmat{ \text{\bf T}}
\def\ti{ t}
\def\tf{\bar t}
\def\xi{\x}

\def\tend{{t_\text{\tiny L}}}
\def\tform{{t_\text{f}}}
\def\tdecoh{t_\text{d}}

\def\tbr{t_\text{br}}
\def\oDC{\omega_\text{\tiny DC}}
\def\tquant{t_\text{quant}}

\def\Gc{{\cal G}}

\def\Mc{{\cal M}}
\def\Jc{{\cal J}}

\def\Dc{{\cal D}}
\def\Kc{{\cal K}}
\def\pT{p_{\scriptscriptstyle T}}

\def\Qc{\mathscr{Q}}
\def\Pc{\mathscr{P}}
\def\Pct{\mathscr{\tilde P}}
\def\Sc{\mathscr{S}}
\def\Cc{\mathscr{C}}

\def\Jc{\mathscr{J}}

\def\dd{\text{d}}

\def\dd{\text{d}}

\def\abar{\bar\alpha}

\def\abar{\bar\alpha}

\begin{document}

\title{Higher-order corrections to heavy-quark jet quenching}

\author{Boris Blok}
\email{blok@ph.technion.ac.il}
\affiliation{Department of Physics, Technion - Israel Institute of Technology, Haifa, Israel}

\author{Konrad Tywoniuk}
\email{konrad.tywoniuk@uib.no}
\affiliation{Department of Physics and Technology, University of Bergen, 5020 Bergen, Norway}

\date{\today}

\begin{abstract}
We calculate higher-order corrections to the quenching factor of heavy-quark jets due to hard, in-medium splittings in the framework of the BDMPS-Z formalism.
These corrections turn out to be sensitive to a single mass-scale $m_\ast = (\hat q L)^{\onehalf}$, where $\hat q$ is the medium transport coefficient and $L$ the path length, and allow to draw a distinction between the way light, with $m < m_\ast$ (in contrast to massless $m=0$), and genuinely heavy, with $m > m_\ast$, quark jets are quenched in the medium. We show that the corrections to the quenching factor at high energies are double-logarithmic and qualitatively of the same order as for the massless quark jet.
\end{abstract}

\maketitle

\section{Introduction}
\label{sec:intro}

Jets are formed in the process of soft and collinear QCD radiation that results in a spray of collimated hadrons and energy deposition in the detector \cite{Dokshitzer:1991wu}. In heavy-ion collisions, partons traverse a hot and dense nuclear medium that leaves an imprint on the subsequent jet formation, for reviews see \cite{dEnterria:2009am,Mehtar-Tani:2013pia,Blaizot:2015lma}. Currently it is widely accepted that the BDMPS-Z formalism of radiative energy loss \cite{Baier:1996kr,Baier:1996sk,Zakharov:1997uu} describes the propagation and multiple scattering of quark and gluon jets in the nuclear QCD medium that is produced in heavy ion collisions at LHC, for a review see Ref.~\cite{Baier:2000mf}.

Jet quenching is a multi-scale problem. Even for massless partons, there is an convoluted interplay between the intrinsic jet scales, such as the mass of the jet, and the scales of the medium, including typically the medium transport coefficient and the medium size. For quarks, the non-zero mass introduces another scale. It is well known that the collinear divergence is regulated by the characteristic dead-cone angle
\beq
\label{eq:dead-cone-angle}
\Tdc = \frac{m}{E} \,.
\eeq
where $m$ is the mass of the heavy quark and $E\equiv p_T$ is the jet energy. As a consequence of the strong suppression of gluon radiation inside the dead-cone angle, heavy-quark jets fragment differently from jets originating from their massless counterparts or from gluons \cite{Dokshitzer:1991wu,Dokshitzer:1982xr,Dokshitzer:1991fd,Dokshitzer:1995ev}.

Radiative energy loss was calculated by BDMPS-Z for massless partons \cite{Baier:1996kr,Baier:1996sk,Zakharov:1997uu}. It was first pointed out in Ref.~\cite{Dokshitzer:2001zm} that the quenching of massive quarks would be different from massless ones because of the dead-cone effect. The resulting restriction of the phase space for radiation, and hence energy loss, leads to a systematically smaller suppression of single-inclusive hadron spectra the larger the mass of the constituent quarks. This was followed up by a more thorough analysis in \cite{Armesto:2003jh,Armesto:2005iq}, where it was shown that the heavy quark quenching factors get further corrections when the correct phase  space constraints are taken into account. For results within the limit of dilute media, see also \cite{Djordjevic:2003zk,Zhang:2018nie}. In summary, based on radiative processes alone one expects a smaller rate of emissions off massive quarks compared to massless ones, that brings about a mass-hierarchy of the suppression. In contrast, low-$\pT$ heavy mesons have a similar modification as the pions \cite{Sharma:2011zz,Wang:2017vrn,CMS:2017dec}. This has prompted many investigations of additional elastic energy loss processes, for a review see \cite{Qin:2015srf}. It is however worth keeping in mind that the final suppression of heavy mesons and heavy flavor jets depends also on the details of the partonic cross sections and the problem is still an open one.

While most of the previous contributions has focussed on small $\pT$, where the cross sections are the largest, we will mainly focus on the genuinely high-$\pT$ regime where perturbative corrections play a crucial role. This regime is within the reach of the experiments at LHC, see e.g. \cite{Chatrchyan:2013exa,CMS:2017dec} . 
Recently higher-order corrections to the quenching of independent, massless quark/gluon jets were calculated  \cite{Mehtar-Tani:2017ypq,Mehtar-Tani:2017web}. The results demonstrate how these contribution lead to the enhanced quenching of massless quark/gluon jets as compared to single partons. The role of in-medium jet splittings and their color coherence properties has also been emphasized in other contexts, see e.g. \cite{Casalderrey-Solana:2017mjg,Caucal:2018dla}.
Consequently, it will be of interest to extend the previous efforts to include mass effects. In this work, we consider higher-order corrections to the quenching of a heavy-quark jet, i.e. a jet formed as a result of the fragmentation of a leading massive quark.

Our main result is that higher-order corrections lead to an enhanced suppression for heavy-quark jets relative to the leading BDMPS-Z result, corresponding to the quenching of a single parton. The magnitude is determined by the phase space available for the radiation of hard gluons within the jet, and is of similar magnitude as for quark/gluon jets in general. However, due to the restricted phase space determined by the dead-cone \eqref{eq:dead-cone-angle}, the mass sets the scale where significant deviations between massive and massless jets can be observed. We identify a critical mass scale that permits to observe such discriminating features in the high-$\pT$ regime.

The paper is organized in the following way. In \autoref{sec:qw-1}, we introduce the generalized quenching weight and discuss its expansion in terms of the strong-coupling constant. We calculate the radiative energy loss due to multiple, soft BDMPS-Z radiation off a single heavy quark and a heavy quark-gluon dipole in \autoref{sec:computing}, and obtain the evolution equations and expressions for related quenching factors. The details of the calculations of the associated spectra and rates are given in Appendices~\ref{sec:antenna-spectrum} and \ref{sec:direct-interference-rates}, respectively, where the basic formulae for the interference contributions to antenna radiation are derived in detail. In \autoref{sec:hq-collimator}, we finally map out the logarithmic phase space for higher-order corrections and present explicit expressions for the collimator function of heavy-quark jet together with numerical results. We summarize our results and give a brief outlook in  \autoref{sec:conclusions}.

\section{Generalized quenching weight}
\label{sec:qw-1}
   
Assuming small energy losses in the medium, $\epsilon \ll \pT$, and accounting for a steeply falling hard spectrum, the spectrum of heavy-quark jets in heavy-ion collisions can be written as
\beq
\frac{\dd \sigma}{\dd \pT^2} = \int_0^\infty \dd \epsilon \, \Pc (\epsilon,L \vert m) \left.\frac{\dd \sigma_0}{\dd q_{\tiny T}^2}\right\vert_{q_{\tiny T} = \pT + \epsilon} \simeq \frac{\dd \sigma_0}{\dd \pT^2} \mathscr{Q}(\pT)  \,.
\label{slon1}
\eeq
where $\dd \sigma_0/\dd \pT^2$ is the Born-level jet production cross section,  $ \Pc (\epsilon,L \vert m)$ is an energy-loss probability distribution associated with a massive particle and $L$ is the medium length (below we shall suppress the arguments $L$ and $m$, unless it is unclear from the context). The jet suppression factor $\Qc(\pT)$, introduced in the second step, is the Laplace transform of the energy loss distribution $\Pc (\epsilon)$, i.e. $\Qc(\pT) \equiv \Pct(n/\pT) = \int_0^\infty \dd \epsilon \, \rme^{ - n \epsilon /\pT} \Pc (\epsilon)$ \footnote{Assuming that $\dd \sigma$ we have only accounted for the first term in the expansion $(1+x)^{-n} \approx \rme^{-nx} (1+ n x^2/2 + \ldots)$.}, where the effective power of the steeply falling spectrum is $n = \frac{\dd}{\dd \ln \pT} \ln \frac{\dd \sigma_0}{\dd \pT^2}$ \cite{Baier:2001yt,Dokshitzer:2001zm}. The jet suppression factor permits an expansion in the strong-coupling constant that accounts for the energy loss of in-medium jet splittings,
\beq
\label{eq:quenching-factor-expansion}
\Qc(\pT) = \Qc^{(0)}(\pT) + \Qc^{(1)}(\pT) + \mathcal{O}(\alpha_s^2) \,.
\eeq
The first term in the expansion is the quenching of the jet total charge which, for a heavy-quark initiator, is given by $\Qc^{(0)}(\pT) = \Pct_q(n/\pT)$. This distribution is dominated by soft gluon radiation that transfers energy from the jet axis to large angles.

The resummation of higher-order terms leads to an additional suppression factor which was referred to as the ``collimator'' function in Ref.~\cite{Mehtar-Tani:2017web}. These corrections correspond to the energy-loss of composite, partonic systems created inside the medium during the jet formation. Hard splittings in the jet cone can be described by vacuum splitting functions. Hence the next-to-leading correction to the jet quenching factor, that involves the (real and virtual) emission and subsequent quenching of an additional gluon \cite{Mehtar-Tani:2017web}, takes the form
\beq
\label{eq:quenching-fac-higher-order}
\Qc^{(1)}(\pT) = \int^{R^2}_0 \frac{\theta^2 \dd \theta^2}{(\theta^2 + \Tdc^2)^2} \int_0^1 \dd z \,\frac{\alpha_s}{2\pi} P_{gq}(z) \Theta(\tform \ll L) \, \big[ \Qc_{gq}(\theta,\pT|m) -\Qc_{q}(\pT|m) \big] \,,
\eeq
where $P_{gq}(z)$ is the Altarelli-Parisi splitting function and $\Qc_{gq}(\theta,\pT)$ is the quenching factor of a composite quark-gluon system propagating in the medium \cite{Mehtar-Tani:2017ypq}. This equation holds whenever the splitting takes place early in the medium. This enforces the formation time, $\tform = 2 /[z(1-z) E \theta^2 + \frac{z}{1-z}E \Tdc^2]$, to be short compared to that of any process in the medium, in particular the medium length $L$. As will be discussed in more detail later, an important time-scale is the so called decoherence time $\tdecoh$ which corresponds to the time when a dipole of size $x_\perp \sim \theta t$, characterized by its opening angle $\theta$, is resolved by medium fluctuations. The characteristic wave-length of the latter can be estimated via diffusive broadening as $\lambda_\perp \sim (\hat q t)^{-1/2}$. The two length-scales become of the same order at $\tdecoh \sim (\hat q \theta^2)^{-1/3}$. In this limit, $\tform \ll \tdecoh \ll L$, the splitting process completely factorizes out on the level of the cross section \cite{Casalderrey-Solana:2015bww} and effectively forms a color-charged antenna. This composite system undergoes further medium-induced radiation in the medium that turns out to be sensitive to its opening angle \cite{Mehtar-Tani:2017ypq}.  However,  there can also be strong cancellations between the two quenching factors in the squared brackets in \eqref{eq:quenching-fac-higher-order} for small-angle emissions, when $\tdecoh > L$ or $\theta < \theta_c$, where $\theta_c \sim (\hat q L^3)^{-1/2}$, due to interference effects. 

Higher-order corrections naturally follow a similar logic, becoming sensitive to more complicated radiation patterns.
In the large-$N_c$ limit the picture is simplified further, since a jet in this case can be decomposed into a set of mutually independent color-singlet dipoles, whose radiation is added to that of the total charge radiation \cite{Bassetto:1984ik}. The quenching of the total charge can therefore be factorized out, and the total quenching factor becomes,
\beq
\label{eq:collimator-definition}
\Qc(\pT | m) = \Qc_q(\pT|m)\, \Cc(\pT, R|m) \,,
\eeq
where $\Cc(\pT,R|m)$ is the collimator function that accounts for the quenching of higher-order (real and virtual) jet emissions.
The resummation of such emissions takes, in the general case, the form of a non-linear evolution equation for the collimator function but, in the limit of strong quenching, one can neglect all real emissions and resum the virtual terms, i.e. the second term in the squared brackets of \eqn{eq:quenching-fac-higher-order}. We will discuss the collimator in more detail in \autoref{sec:hq-collimator}. In the remaining part of the paper, we will describe the radiative quenching of a heavy-quark system and identify the relevant time-scales that play a role in this problem in order to compute and resum these corrections. We focus on the high-$\pT$ regime and (relatively) large quark masses, where elastic energy losses, see e.g. \cite{Qin:2015srf}, can be neglected. Our results at low-$\pT$ are therefore not completely realistic. However, we emphasize that for \emph{genuinely} heavy quarks the high-$\pT$ regime (meaning $\pT \gtrsim 20-50$ GeV, see below for more details) is relevant for the phenomenology of heavy-quark jets.

\section{Computing the quenching factors}
\label{sec:computing}

In this section we compute the quenching weights, that is energy loss probability distributions that resum multiple soft, gluon radiation responsible for transporting energy from the leading particle to large angles. As mentioned in the Introduction, it will be convenient to work directly in Laplace space, defined as
\beq
\Pc(\epsilon,L| m) = \int_C \frac{\dd \nu}{2 \pi i} \, \Pct(\nu,L |m) \rme^{\nu \epsilon} \,,
\eeq
where the contour $C$ runs parallel to the imaginary axis in the complex-$\nu$ plane, $\rmR \nu =$ const., to the right of any singularity of $\Pct( \nu,L|m)$.  To recap, $\Pc(\epsilon,L|m)$ acts as a probability distribution for radiating gluons that in total carry an energy $\epsilon$ off a particle with mass $m$ after propagating through a medium of length $L$, and the quenching factor $\Qc(\pT) = \Pct(n/\pT, L|m)$. It will be convenient to define a ``regularized'' splitting rate,
\beq
\gamma_{ij}(\nu,t) = \int_0^\infty \dd \omega\, \left(\rme^{- \nu\omega} -1 \right) \Gamma_{ij}(\omega,t) \,,
\eeq
where $\gamma_i (\nu, t) \equiv \gamma_{ii}(\nu,t)$ and we have already anticipated the possibility of interference contributions between two different particles labeled ``$i$'' and ``$j$'' that refer to quarks, antiquarks or gluons. Here $\Gamma_{ij}(\omega,t)$ is the rate of (interference) emissions in the medium, where the soft gluon is emitted by a parton i and absorbed by the parton j in the complex conjugate amplitude. We derive the generic interference spectrum off a color-charged antenna in Appendix~\ref{sec:antenna-spectrum}, and derive concrete expressions for the direct and interference rates in Appendix~\ref{sec:direct-interference-rates} within the multiple-soft scattering approximation.

\subsection{Quenching of a single parton}
\label{eq:single-parton-quenching} 

Let us start by considering a single propagating particle. Medium interactions can enhance the probability of gluon emissions.
In Laplace space, the resummation of soft, medium-induced gluons takes the form of a rate equation,
\beq
\label{eq:single-particle-evolution}
\partial_t  \Pct_i(\nu,t) = \gamma_i(\nu,t)  \Pct_i(\nu,t) \,,
\eeq
with initial condition $\Pct(\nu,0) =1$, whose solution is simply given by $\Pct_i(\nu,L) = \rme^{\int_0^L \dd t \, \gamma_i(\nu,t)}$. For a massive quark, the emission rate of soft gluons was derived in \eqn{eq:massive-spectrum-direct-scaling}, and reads
\beq
\label{eq:massive-quark-rate-2}
\Gamma_{q}(\omega,t \vert m)=\abar\, \Theta^2 \left[ - \rmI \,\psi_0\left(-\frac{1+i}{4} \zeta^{3/2} \right) - \zeta^{-3/2} - \frac{3}{4}\pi \right] \,,\label{LQ6}
\eeq
where $\psi_0(x)$ is the digamma function, $\abar \equiv \alpha_s C_F/\pi$ and the expression in the squared brackets is a function of the scaling variable $\zeta \equiv \omega/\oDC$, where 
\beq
\oDC = \big(\hat q \big/\Tdc^4 \big)^{1/3} = \big( \hat q\, \pT^4/m^4 \big)^{1/3} \,.
\eeq
The expression \eqref{eq:massive-quark-rate-2} is valid only for $\zeta < 1$.  It turns out the spectrum is strongly suppressed at $\omega > \oDC$, where we also observe negative contributions owing to the treatment of the high-energy behavior which goes beyond the leading-logarithmic accuracy of our calculation \footnote{All results for the medium-induced spectra and rates hold in the soft limit where we have explicitly subtracted the vacuum component. Negative contributions in these quantities therefore indicate an over-subtraction at larger gluon energies.}. In order to avoid these unphysical contributions, and retain the information about the physical scales, we approximate the rate by
\beq
\Gamma_{q}(\omega,t \vert m) \approx \abar\sqrt{\hat q / \omega^3} \, \Theta(\oDC - \omega) \,.\label{LQ4}
\eeq
In Laplace space, this becomes
\begin{align}
\gamma_q(\nu,t \vert m) &= 2\abar \sqrt{\frac{\hat q}{\oDC}} \big[1 - \rme^{-\nu \oDC} - \sqrt{\pi \nu \oDC}  \,\text{erf}\left(\sqrt{\nu \oDC} \right)\big] \,, \\
\label{eq:HQ}
&\approx -2 \abar \sqrt{\hat q} \big( \sqrt{\pi \nu } - \sqrt{\oDC}\big) \qquad\qquad \text{for } \nu^{-1} \lesssim \oDC \,,
\end{align}
where $\text{erf}(x)$ is the error function. This is qualitatively similar to what is obtained in Ref.~\cite{Dokshitzer:2001zm}, although the precise form of the cut-off at $\oDC$ determines a numerical constant in front of the second term in the brackets (according to the authors of Ref.~\cite{Dokshitzer:2001zm}, this factor is $\sim 1.5$). The corresponding rates for a massless quark is found by taking $m \to 0$ and for gluon by futher replacing the color factor $C_F \to N_c$, e.g.
\beq 
\label{eq:LQ1}
\Gamma_q(\omega,t \vert m=0) = \abar \sqrt{\hat q/\omega^3} \qquad \text{and} \qquad \gamma_g(\nu,t \vert m =0) = -2 \abar \sqrt{\pi \hat q \nu} \,,
\eeq
for massless quarks. Neglecting corrections $\mathcal{O}(1/N_c)$, which shortly will be further motivated, we find that the gluon rates by $\Gamma_g=2\Gamma_q$ and $\gamma_g=2\gamma_q$.

While these rates are time-independent in the limit of soft gluon emissions, this is violated at large energies. This approximation breaks down for emissions with formation times of the order of the medium length, corresponding to a critical energy $\omega_c \sim \hat q L^2$ that brings about  a power-like cut-off of the spectrum and, therefore, the rate as well. The constraint from the dead-cone angle is stronger than this absolute limit whenever $\oDC < \omega_c$ which, in turn, implies that $\Tdc > \theta_c$. This marks the regime where the mass of the quark should start affecting the general properties of radiative energy loss that is dominated by LPM interference effects.

\subsection{Quenching of a two-parton system (color-charged dipole)} 

Let us now turn to the higher-order corrections to this picture, that arise from a quark-gluon antenna propagating in dense QCD media. Considering for the moment the energy loss of a quark-gluon dipole that is formed quasi-instantaneously after the hard vertex, in Laplace space the joint energy loss distribution factorizes in the large-$N_c$ limit into the product of energy loss off a total charge (triplet)
and a color-singlet dipole, 
\beq
\label{eq:slon2}
\Pct_{gq}(\nu,t | m) = \Pct_{q_0}(\nu,t) \Pct_{q_1 q_2}(\nu,t | m) \,,
\eeq
where $ \Pct_{q_0}(\nu,t) \equiv  \Pct_{q_0}(\nu,t | m=0) $ and we have decomposed the gluon into a quark-antiquark pair $g=(q_0,q_1)$ (where $q_1$ is an antiquark). Recall that only the quark that forms part of the dipole $q=q_2$ is massive. Note that the quenching of the total (quark) charge is not sensitive to the mass of the initial particle, since it is inherited from the radiated gluon. Instead, the mass controls the energy loss of the additional irreducible singlet $\Pct_{q_1 q_2}(\nu,t | m) \equiv \Pct_\text{sing}(\nu,t | m)$. We will confirm below that, in the completely decoherent limit, the mass will be associated with the total color charge, as expected.

The two factors in \eqref{eq:slon2}  satisfy two separate evolution equations. First, the single-particle quenching is given by \eqn{eq:single-particle-evolution}, where the splitting rate $\gamma_q$ is explicitly given by \eqref{eq:LQ1}, keeping in mind that this fictitious quark is massless. The singlet, dipole quenching weight is determined by solving the differential equation \cite{Mehtar-Tani:2017ypq}
\beq
\label{eq:two-particle-evolution}
\partial_t \Pct_\text{sing}(\nu,t | m) = \gamma_\text{dir}(\nu,t) \Pct_1(\nu,t)  \Pct_{2}(\nu,t |m)  + \gamma_\text{int}(\nu,t) \Sc_2(t) \,.
\eeq 
The initial condition at $t=0$ (corresponding to the time when the antenna was formed in the medium) is again trivial, $\Pct_\text{sing}(\nu,0|m)=1$. The direct and interference rates were derived in Appendix~\ref{sec:direct-interference-rates}, and are given by
\begin{align}
\gamma_\text{dir} (\nu,t) &= \gamma_{1}(\nu,t) + \gamma_2 (\nu,t) \,, \\
\gamma_\text{int}(\nu, t) &= \gamma_{12}(\nu,t) + \gamma_{21}(\nu,t) \,.
\end{align}
The two direct terms correspond to emissions off the two legs, and similarly the interference terms correspond to emitting a gluon from one leg and ``absorbing'' it (in the complex conjugate amplitude) on the  other (see \autoref{fig:diagram} for details). For the singlet dipole we have $\gamma_\text{dir}(\nu,t) = 2 \gamma_q(\nu, t)$, since $\gamma_1(\nu,t) = \gamma_2(\nu,t)\equiv \gamma_q(\nu,t)$. 

Note that \eqn{eq:two-particle-evolution} contains a inhomogeneous term arising from the possibility of interferences between the dipole consituents. The interference spectra have a more complex structure since they involve both color and quantum decoherence processes \cite{MehtarTani:2011tz,CasalderreySolana:2011rz,MehtarTani:2011gf}. The interference spectrum associated with a massive dipole is given explicitly in \eqn{eq:final-spectrum-1a}, see also \cite{Calvo:2014cba}, and evaluated in the multiple-soft scattering approximation in Eqs.~\eqref{eq:interference-approximations-1} and \eqref{eq:interference-approximations-2}. As discussed further in Appendix~\ref{sec:direct-interference-rates}, color decoherence is related to the survival probability of a color-singlet dipole and is explicitly factorized out in the so-called decoherence parameter $\Sc_2(t)$ in \eqref{eq:two-particle-evolution}. This factor is responsible for the previously introduced time scale $\tdecoh \sim (\hat q \theta^2)^{-1/3}$. At late times, $t > \tdecoh$, the dipole decoheres and the particles can radiate independently. Conversely, a coherent splitting corresponds to the situation when $\tdecoh > L$ and the pair remains coherent during the passage through the medium. This applies to the regime of small angles, $\theta < \theta_c$ where $\theta_c \sim (\hat q L^3)^{-1/2}$.

In our formulation, the medium-induced interference rates (in energy-space) are themselves suppressed at a time-scale $\tquant \sim (\theta^2 \omega)^{-1}$. This scale can nevertheless be neglected by noting that $\tquant \sim (\theta_\text{br}(\omega)/\theta)^{4/3} \tdecoh$, where $\theta_\text{br}(\omega) = (\hat q/\omega^3)^{1/4}$ is the branching angle of medium-induced gluons. Since energy loss is governed by soft gluons, $\omega \sim \abar^2 \omega_c$, that parametrically go to large angles, in particular out of the jet cone $\theta_\text{br}(\omega)\sim \abar^{-3/2} \theta_c > R > \theta$, this implies that $\tdecoh < \tquant$ \cite{CasalderreySolana:2011rz,Mehtar-Tani:2017ypq}. Since the multiplicity of hard gluons is small, $N(\omega > \abar^2 \omega_c) \sim \mathcal{O}(\alpha_s)$ and correspondingly their contribution to energy-loss is small \cite{Baier:2001yt} we will neglect such quantum effects in the following. Hence, for our purposes, i.e. at times $t < \tdecoh < \tquant$, the interference rate is approximated as $\gamma_\text{int} (\nu,t) \approx - \gamma_\text{dir}(\nu,t)$, see Eqs.~\eqref{eq:interference-laplace-1} and \eqref{eq:interference-laplace-2}. This property is independent of the mass.

The solution to the rate equation can  be written symbolically as
\begin{align}
\Pct_\text{sing} (\nu,L|m)  &= \Pct_1(\nu,L) \Pct_2(\nu,L|m) \nn
&+ \int_0^L \dd t \,  \Pct_1(\nu,L-t) \Pct_2(\nu,L-t | m) \gamma_\text{int}(\nu,t) \Sc_2(t) \,.\label{basic}
\end{align}
The extension of the time-integral of the second term is limited by the shortest time-scale where interferences are suppressed. In the leading-logarithmic approximation
it is sufficient to consider only large-angle radiation where, parametrically, the energy radiated via medium-induced gluons leave the jet cone.
In this case the integral is limited by $t < \tdecoh$, as discussed above, and the singlet distribution can then be approximated by 
\beq
\label{eq:base}
\Pct_\text{sing}(\nu,L|m) \approx \Pct_{q_1}(\nu,L - \tdecoh) \Pct_{q_2}(\nu,L-\tdecoh| m) \,,
\eeq
where we have reinstated the mass dependence. Hence, the decoherence time acts as a ``delay'' for when energy loss processes start affecting the irreducible dipole and, in the limit $\tdecoh \ll L$, the dipole constituents decohere early in the medium and lose energy independently along the whole medium length.

To summarize, the delay effect is strictly associated with the color dynamics of the dipole and, since this involves the shortest relevant time-scale, does not depend on the mass of the constituents. It might, at first look, seem strange that the mass-effect on quenching is delayed although it is intimately linked with the quark-initiator and, hence, the total charge. For instance, considering long decoherence times, $\tdecoh \geq L$, applying to small-angle emissions $\theta \leq \theta_c$, the color-charged antenna is quenched as a \emph{massless} quark, rather than a \emph{massive} one. This effect gives rise to a mismatch between real and virtual emissions at small angles. This turns nevertheless out to be a sub-leading effect, see \eqn{eq:collimator-massive-2}.

\section{Heavy-quark collimator function}
\label{sec:hq-collimator}

\subsection{Quenching of total charge}

The first term in the expansion in \eqref{eq:quenching-factor-expansion} corresponds to the quenching of a single, massive quark. After implementing the result in \eqref{eq:HQ}, we find that 
\beq
\label{eq:heavy-quark-total-quench}
\Qc^{(0)}(\pT|m) \equiv \Qc_q(\pT|m) = \exp \left[- 2 \abar L \left(\frac{\pi \hat q n}{\pT} \right)^{1/2} \right] \exp\left[ 2 \abar L \left(\frac{\hat q m^2}{\pT^2} \right)^{1/3} \right] \,,
\eeq
where the first term corresponds to the quenching of a massless color parton, while the second is a mass-dependent enhancement factor. The mass-independent term 
implies that the regime of strong quenching of massless quarks, $- \ln \Qc^{(0)} \sim \mathcal{O}(1)$, is given by 
\beq
\pT < n \omega_s \,,
\eeq 
up to numerical factors, where $\omega_s \sim \abar^2 \hat q L^2$ is a soft scale corresponding to large multiplicity of medium-induced emissions, $N(\omega < \omega_s) > \mathcal{O}(1)$. We can rewrite the quenching factor as $\Qc^{(0)}(\pT|m=0) \approx \exp \big[ - \sqrt{\pi} N(\pT/n) \big]$ \cite{Baier:2001yt}, which is
interpreted as a Sudakov suppression factor for medium-induced gluons with energies $\omega> \pT/n$.

The regime with an  additional strong enhancement of heavy compared to massless quarks arises for 
\beq
\label{eq:hq-strong-quenching}
\pT < m  \theta_s^{-1} \sim \abar^{3/2} m (\hat q L^3)^{1/2} \,,
\eeq
where $\theta_s \sim \theta_\text{br}(\omega_s) \sim \abar^{-3/2}\theta_c$. This condition is equivalent to demanding that $\theta_s < \Tdc$, which implies that the regime of multiple, soft gluon emissions is cut off by the dead-cone angle. So there exists a regime of strong massless-quark quenching with additional strong effects from heavy-quark quenching whenever for masses $m < n \abar^{1/2} Q_s$ where $Q_s \sim \left(\hat q L \right)^{1/2}$. To limit the scope of our qualitative analysis, we will assume that the index of the steeply falling spectrum $n$, combined with the medium parameters $\hat q$ and $L$, is large enough to work in the regime of strong quenching effects.

\subsection{Scale analysis}

Let us now turn to the next-to-leading correction \eqref{eq:quenching-fac-higher-order}, coming from the emission of an additional hard gluon off the initiating heavy quark early in the medium.
We will currently focus on the leading-logarithmic contributions, leaving an analysis of sub-leading logarithmic contributions for the future. In this context we only consider strong ordering of scales
and we will therefore not typically keep track of numerical factors that are anyway beyond the precision of this analysis. We will also work in the large-$N_c$ limit, where we can exploit the factorization of the color-charged dipoles, as in \eqn{eq:slon2}.

\subsubsection{Massless quarks}

Before turning to effects related to the mass of the jet particles, let us summarize the scale analysis for massless partons. In terms of angles, we have several characteristic scales:
the jet radius $R$, the minimal medium resolution angle $\theta_c \equiv \theta_\text{br}(\omega_c)$ and the typical angle for soft gluon emissions $\theta_s \equiv \theta_\text{br}(\omega_s)$. Note that the two medium scales are parametrically separated by the smallness of the coupling constant, $\theta_s \sim \abar^{-3/2} \theta_c $. 

If $R > \theta_s$ the energy loss of jet is small since all radiated BDMPS gluons remain inside the jet. On the other hand, if $\theta_c > R$, i.e. the jet angle is less than decoherence angle, the propagation of the jet is not influenced by subjet structure, and it is equivalent to propagation of a total color charge, i.e. one parton through the medium. The typical ordering that interests us, where the quenching could be substantial and where higher-order effects are non-trivial, is therefore
\beq
\theta_s > R > \theta_c \,.
\eeq
In what remains, we will assume that this hierarchy holds and, besides, that it is also the phenomenologically most relevant one. 

However, note that the minimal angle $\theta_c$ only is relevant for high-$\pT$ jets, $\pT > \omega_c$. Conversely, for $\pT < \omega_c$ the decoherence time is necessarily always shorter than the medium length, $\tdecoh < L$. In this case there is still the possibility for a regime of short formation times, $\tform < \tdecoh$, but in this case this condition implies that $\theta > \theta_d$, where
\beq
\label{eq:thetad}
\theta_d \sim \left(\frac{\hat q }{\pT^3} \right)^{1/4} \,.
\eeq
For massless quarks, this regime is double-logarithmic in the jet scale \cite{Mehtar-Tani:2017web}, see below, but the $\pT$-range is automatically limited by $\omega_c$. The window for a regime of short formation times closes whenever $\theta_d = R$, or $\pT \sim (\hat q/ R^4)^{1/3}$. In the following, we will therefore distinguish between high-$\pT$ (with $\pT > \omega_c$) and low-$\pT$ (with $\pT < \omega_c$) jets. 

\subsubsection{Massive quarks}

For massive quarks, the dead-cone angle (energy, etc.) introduces another physical scale to the problem. For a finite dead-cone, QCD radiation is no longer genuinely collinearly enhanced which necessitates a scale-dependent scheme to properly include mass-effects for resummed observables, see e.g. \cite{Dokshitzer:1991wu}. We will only stick to the leading-logarithmic approximation and only consider emissions $\theta > \Tdc$, i.e. $\theta^2/(\theta^2+ \Tdc^2)^2 \approx \theta^{-2} \Theta(\theta-\Tdc)$ in \eqref{eq:quenching-fac-higher-order}. Hence, for $R \sim \Tdc$ the heavy-quark jet only contains a single quark.

Comparing the mass scale to other relevant medium scales, in particular comparing $\Theta$ and $\theta_c$, can become involved because of the $\pT$-dependence of the former.
In order to organize the discussion, it will be useful to introduce a critical value of the mass, namely
\beq
m_\ast \equiv (\hat q L)^\onehalf \,.
\eeq
From now on we will call quarks with $ m > m_\ast$ genuinely \emph{heavy}, and quarks with $m < m_\ast$ for \emph{light} (in contrast to massless, $m = 0$). We have also summarized the discussion about the relevant scales in \autoref{tab:scalesummary}.

\begin{table}[t]
\centering
\begin{tabular}{c|c|c|c}
Quark mass & Distinctive angle & Critical jet $\pT$ & Critical parton $\pT$ \\
\hline
$m < m_\ast$ & $ (\hat q/\pT^3)^{1/4}$ & $m^4/\hat q$  &$\abar^{3/2} m (\hat q L^3)^{1/2}$ \\
$m > m_\ast$ & $ (\hat q L^3)^{-1/2}$ & $ m (\hat q L^3)^{1/2}$ & $\abar^{3/2} m (\hat q L^3)^{1/2}$
\end{tabular}
\caption{Summary of the scales for light and heavy quarks.}
\label{tab:scalesummary}
\end{table}

For \emph{heavy} quarks, the dead-cone angle becomes comparable to the coherence angle, $\Tdc = \theta_c$, at large-$\pT$, i.e. $\pT > \omega_c$. Rewriting the same condition, this happens at a critical energy $\pT = m \theta_c^{-1} \sim m (\hat q L^3)^{1/2}$. Hence, we expect the heavy-quark jet quenching to deviate from the light-quark jet quenching at a scale that is parametrically larger, by a factor $\abar^{-3/2}$, than the soft scale identified for the quenching of the total charge, cf. \eqn{eq:hq-strong-quenching}. In other words, while the quenching of a single heavy quark starts deviating from the massless one at relatively low $\pT$ due to the enhancement factor in \eqref{eq:heavy-quark-total-quench}, a jet initiated by a heavy-quark should start deviating from the behavior of a massless quark or gluon jet already in the high-$\pT$ regime, since by definition $m (\hat q L^3)^{1/2} > \omega_c$. Considering high-$\pT$ single-inclusive mesons to be proxies of single-parton dynamics, see e.g. \cite{Arleo:2017ntr}, this analysis therefore predicts a different behavior of heavy-quark jets and heavy-quark mesons over a large range in $\pT$. We will come back to a possible experimental signature for this effect in \autoref{sec:numerics}.

For \emph{light} quarks, $\Tdc \neq \theta_c$ for $\pT \geq \omega_c$ and we have instead to consider the low-$\pT$ regime, i.e. $\pT < \omega_c$. In this regime, the condition $\tform < \tdecoh$ implies that $\theta > \theta_d$ which is estimated in \eqref{eq:thetad}. Therefore $\Tdc = \theta_d$ when $\pT \sim m^4/\hat q$, which can be considered a relatively soft energy-scale (comparing it to the soft scale in \eqref{eq:hq-strong-quenching} gives $m \sim \abar^{1/2}(\hat q L)^{1/2}$ which is compatible with the prior assumption about the smallness of the mass). We conclude therefore that the light-quark jets behave similarly to massless jets, as far as the higher-order corrections go, and start deviating from this behavior only when $\Tdc \sim \theta_s$, where the quenching of the total charge gets suppressed. This follows very closely the trend of single-parton, or single-inclusive meson, quenching.

\begin{figure}
\centering
\includegraphics[width=0.48\columnwidth]{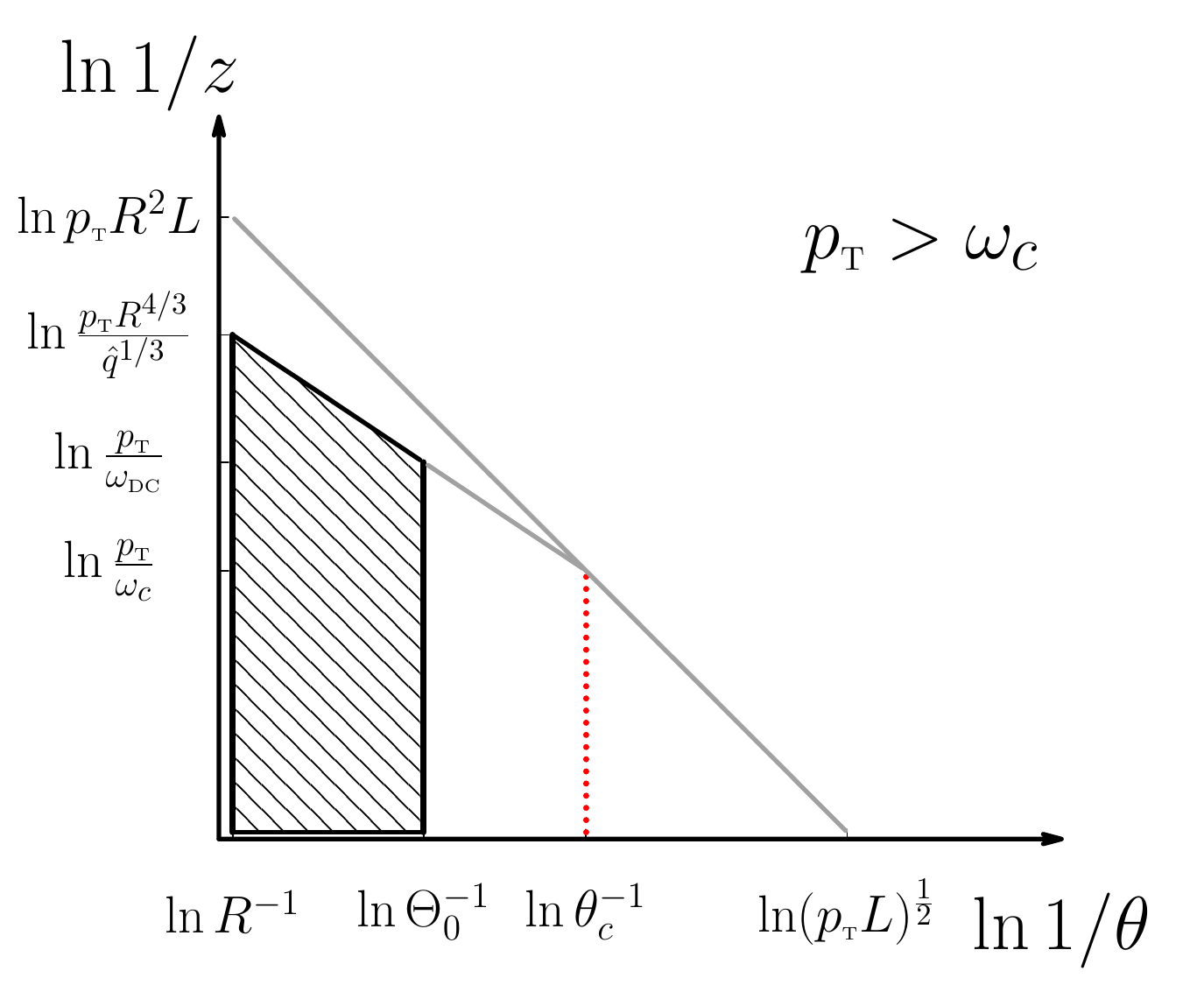}%
\includegraphics[width=0.48\columnwidth]{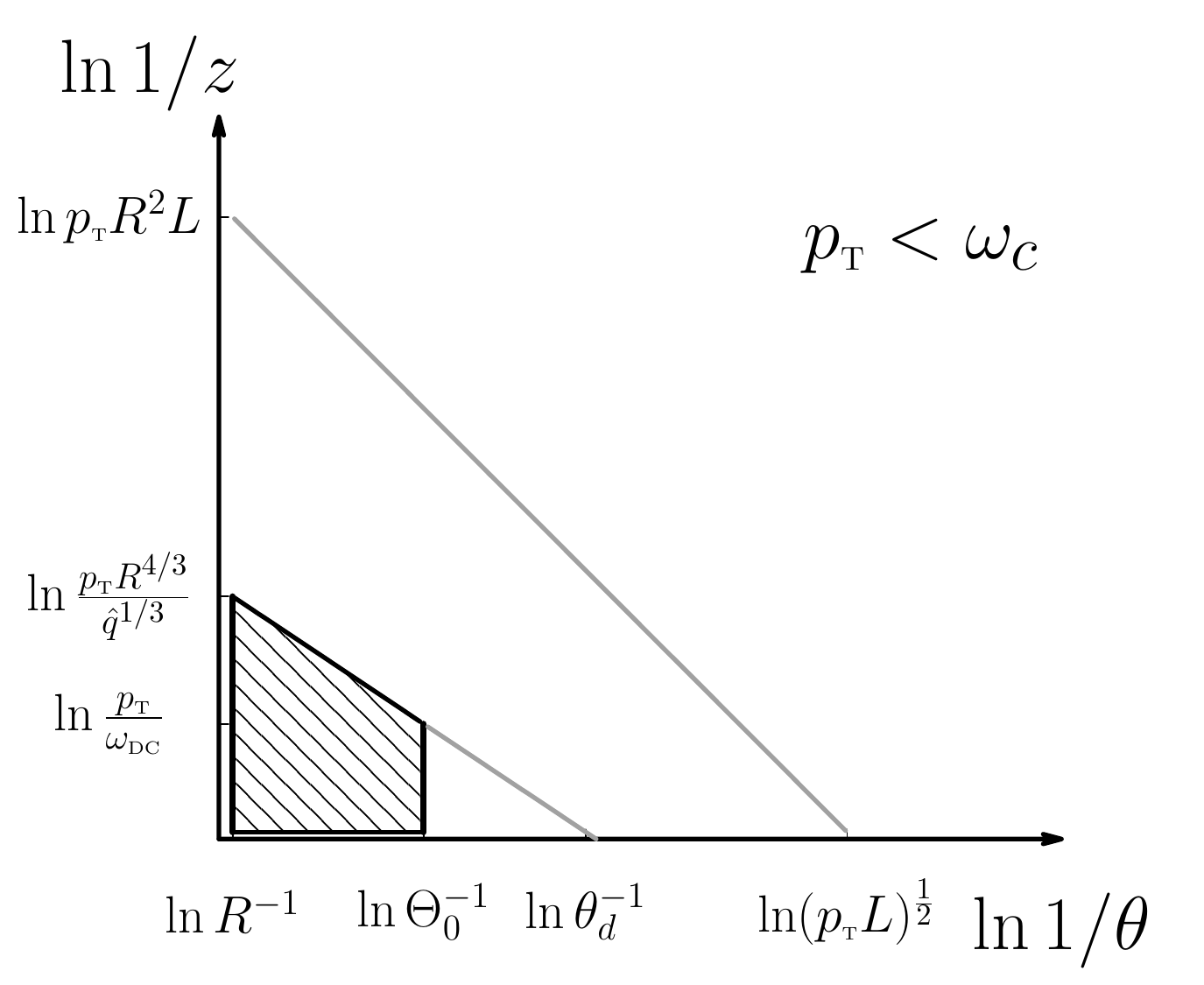}
\caption{Illustration of the DLA phase space for higher-order quenching effects for massive particles, marked by the black, lined area. The two lines correspond to $\tform = L$ (upper line with slope $-2$) and $\tform = \tdecoh$ (lower line with slope $-4/3$).The diagrams are drawn for the two cases: $\pT > \omega_c$ and $R> \Tdc > \theta_c$ (left) and $\pT < \omega_c$ and $R > \Tdc > \theta_d$ (right).}
\label{fig:heavylightjet}
\end{figure}

The corresponding kinematical Lund planes for high- and low-$\pT$ heavy-quark jets are illustrated in \fign{fig:heavylightjet}, where we have spanned the plane in the logarithmic variables $1/z$ and $1/\theta$. At fixed coupling, the plane is equally filled with splittings with probability $2 \abar$, up to a color factor. The two diagonal lines, with slopes $-4/3$ and $-2$, delineate the conditions $\tform = \tdecoh $ and $\tform = L$, respectively. The area between the two lines corresponds to in-medium radiation with $\tdecoh \lesssim \tform < L$, or $k_\perp \lesssim \sqrt{\hat q \omega}$, which is strongly influenced by medium interactions and broadening. 

The high-$\pT$ regime, $\pT > \omega_c$, is plotted on the left side of \fign{fig:heavylightjet}, where we have marked the location of the critical angle $\theta_c$ with a (red) dotted line. Similarly, the dead-cone angle is marked, and corresponds to an energy scale $\oDC$ at $\tform = \tdecoh$. The low-$\pT$ regime, $\pT < \omega_c$ is conversely plotted on the left in \fign{fig:heavylightjet}. One observes immediately that the critical angle $\theta_c$ is replaced by $\theta_d$. In both figures we have assumed that the dead-cone is appreciable, i.e. $\Theta > \theta_c$ for $\pT > \omega_c$ and $\Theta > \theta_d$ for $\pT < \omega_c$, and marked out the phase space available for hard, in-medium splittings of the heavy-quark.

\subsection{Higher-order contributions to quenching}

The analysis in the preceding section allows us to calculate the higher-order contributions to jet quenching. Isolating the quenching of the initiating parton, that corresponds to the total color charge of the jet, into an overall pre-factor, see \eqn{eq:collimator-definition}, these contributions are collected into the collimator function. Using \eqn{eq:quenching-fac-higher-order} and the definition in \eqref{eq:collimator-definition}, we see that the first-order correction the collimator function is 
\begin{align}
\Cc^{(1)}(\pT|m) &\approx 2 \abar \int^R_{\Tdc} \frac{\dd \theta}{\theta} \int_{\min((\hat q/\theta^4)^{1/3}, \,(\theta^2 L)^{-1} )}^{\pT} \frac{\dd \omega}{\omega} \nn
&\times  \left[\frac{\Qc_{q}(\pT)}{\Qc_{q}(\pT|m)} \Qc_q(\pT,L-\tdecoh) \Qc_q(\pT,L-\tdecoh |m) -1 \right] \,,
\end{align}
where we have treated the splitting vertex in the leading-logarithmic approximation and adopted the notations of the previous section.

It is worth pointing out two limits of this equation. For $\tdecoh \ll L$, we can neglect the decoherence times in the real term, i.e. the first term in the squared brackets, to obtain
\beq
\label{eq:collimator-massive-1}
\Cc^{(1)}(\pT|m) \big|_{\tdecoh \ll L} \approx 2 \abar \int^R_{\max(\Tdc, \, \theta_c,\, \theta_d)} \frac{\dd \theta}{\theta} \int_{(\hat q/\theta^4)^{1/3}}^{\pT} \frac{\dd \omega}{\omega} \,  \left[\Qc^2_q(\pT) -1 \right] \,,
\eeq 
which, when taking $m \to 0$, is equal to the contribution of massless quark quenching. When $\Tdc > \max(\theta_c,\theta_d)$ the angular phase space is more restricted for heavy-quark jets, and therefore we expect a relatively smaller impact of the collimator function than in the massless case. Here it is worth pointing out that the quenching factor on the right-hand side of \eqref{eq:collimator-massive-1} arises due to the quenching of the additional (massless) gluon since, at large-$N_c$, $\Qc_q^2(\pT) = \Qc_g(\pT)$, which is a generic property of Sudakov suppression factors.

Before continuing, we point out a new contribution in the small-angle limit in the regime that is unique to massive-quark jets. It appears for $\tdecoh > L $, or $\theta < \theta_c$,  relevant for $\pT > \omega_c$, where $\Qc(\pT, L-\tdecoh |m)= 1$. We are left with
\beq
\label{eq:collimator-massive-2}
\Cc^{(1)}(\pT) \big|_{\tdecoh  > L, \pT >\omega_c} \approx 2 \abar \int^{\theta_c}_{\Theta} \frac{\dd \theta}{\theta} \int_{(\theta^2 L)^{-1} }^{\pT} \frac{\dd \omega}{\omega} \,  \left[\frac{\Qc_q(\pT)}{\Qc_q(\pT|m)} -1 \right] \,.
\eeq
However, $\Tdc < \theta_c$ for $\pT > m \theta_c^{-1}$, which leaves the factor $\Qc(\pT)/\Qc(\pT|m)-1 \gtrsim \abar \mathcal{O}(1)$, and therefore the contribution in this regime is sub-leading $\sim \mathcal{O}(\abar^2)$. We will therefore altogether neglect this regime when working in the leading-logarithmic approximation.

Let us now evaluate the next-to-leading contributions for massless, light and heavy quarks. For completeness, we repeat here the resulting collimator function for massless quarks at first order in $\alpha_s$, that reads \cite{Mehtar-Tani:2017web}
\beq
\label{eq:m1}
\frac{\Cc^{(1)}(\pT|m=0)}{[ \Qc^2_q(\pT) - 1 ]} =
\begin{cases} 
2 \abar \ln \frac{R}{\theta_c} \left( \ln \frac{\pT}{\omega_c} + \frac{2}{3} \ln \frac{R}{\theta_c} \right)   & \text{ for } \pT > \omega_c \,, \\
\frac{3 \abar}{4} \ln^2 \frac{\pT R^{4/3}}{\hat q^{1/3}} &  \text{ for } (\hat q/R^4)^{1/3} < \pT < \omega_c \,.
\end{cases}
\eeq
Turning now to the new results, for light quarks we obtain
\beq
\label{eq:m2}
\frac{\Cc^{(1)}(\pT|m<m_\ast)}{[ \Qc^2_q(\pT) - 1]} =
\begin{cases}
2 \abar \ln \frac{R}{\theta_c} \left( \ln \frac{\pT}{\omega_c} + \frac{2}{3} \ln \frac{R}{\theta_c} \right) & \text{ for } \pT > \omega_c \,, \\
\frac{3 \abar}{4}  \ln^2 \frac{\pT R^{4/3}}{\hat q^{1/3}} &  \text{ for } m^4/\hat q < \pT < \omega_c  \,, \\
 \frac{4 \abar}{3} \ln\frac{\pT R}{m} \left(\ln \frac{\pT R}{m} + \ln \frac{m^2}{(\hat q \pT)^{1/2}} \right) & \text{ for } (\hat q/R^4)^{1/3} < \pT < m^4/\hat q \,,
\end{cases}
\eeq
and for heavy quarks we get instead,
\beq
\label{eq:m3}
\frac{\Cc^{(1)}(\pT|m>m_\ast)}{[ \Qc^2_q(\pT) - 1]} =
\begin{cases}
2 \abar \ln \frac{R}{\theta_c} \left( \ln \frac{\pT}{\omega_c} + \frac{2}{3} \ln \frac{R}{\theta_c} \right) & \text{ for } \pT > m (\hat q L^3)^{1/2} \,, \\
\frac{4 \abar}{3} \ln\frac{\pT R}{m} \left(\ln \frac{\pT R}{m} + \ln \frac{m^2}{(\hat q \pT)^{1/2}} \right) & \text{ for } (\hat q/R^4)^{1/3} < \pT < m (\hat q L^3)^{1/2} \,.
\end{cases}
\eeq
Equations \eqref{eq:m1}--\eqref{eq:m3} are written with logarithmic accuracy, i.e. we neglected all $\mathcal{O}(1)$ numerical factors that enter the arguments of the logarithms. The inclusion of these factors change the scales in the arguments of the logarithms of the order of $1-2$, but does not change any qualitative conclusions  we make.

Let us briefly comment on further contributions to the collimator at higher-order (next-to-next-to-leading, and higher). 
Examining the structure of \eqn{eq:slon2}, one realizes that the dipole that ``contains'' the heavy-quark is distinct from further dipoles in the sense that it is massive while further dipoles, originating from other gluon emissions, are massless. However, as discussed in detail above, this distinction gives rise only to sub-leading corrections and for our purposes, having separated out the specific quenching factor of the originating parton (that also carried the total color charge), it is adequate to treat all dipoles on equal footing.

The problem then reduces to the massless case with a modified phase space, as detailed above. The resummation of higher-order contributions to the collimator involves solving a non-linear evolution equation and was derived in Ref.~\cite{Mehtar-Tani:2017web}. It goes beyond the scope of our investigation to solve this equation here for the massive case. Furthermore, since we are interested in a relatively modest $\pT$ range in order to be sensitive to the dead-cone, the phase space is limited and the first, non-trivial term should provide a good estimate of the effects. Using the same arguments as in \cite{Mehtar-Tani:2017web}, we therefore expect  that the full collimator function is well approximated by
\beq
\label{eq:m4}
\Cc(\pT ) \approx \exp \big[\Cc^{(1)}(\pT) \big] \,.
\eeq
In particular, the strong quenching limit, $\Qc(\pT) \ll 1$, returns the correct exponentiation of the virtual terms and also the fixed point at $\pT \to \infty$, where $\Qc(\pT) \to 1$ and therefore $\Cc(\pT) \to 1$, is reproduced.

\subsection{Numerics}
\label{sec:numerics}

To emphasize the effects of higher-order contributions we propose the following phenomenological quantity,
\beq
\label{eq:osm}
J_\text{\tiny AA}(\pT,R|m) = \frac{R^\text{jet}_\text{\tiny AA}(\pT,R|m)}{R^\text{meson}_\text{\tiny AA}(\pT|m)} \,,
\eeq
which is a ratio of nuclear modification factors of heavy-quark jets to heavy-quarks. Within our approximations, this ratio is simply the collimator function for massless and massive quarks $J_\text{\tiny AA}(\pT,R|m) \approx \Cc(\pT,R|m)$, where we have utilized that ${R^\text{jet}_\text{\tiny AA}(\pT,R|m)} \simeq \Qc(\pT,R |m)$ and $R_\text{\tiny AA}^\text{meson} \simeq \Qc_q(\pT|m)$. The latter can be justified in the sense that, although the quenching factor of a heavy-quark is not a direct measurable, it is closely related to the quenching factor of the corresponding heavy-meson \cite{Arleo:2017ntr}.  The underlying idea builds on the assumption of a similar path-length dependence for the two observables.

\begin{figure}[t]
\centering
\includegraphics[width=0.5\textwidth]{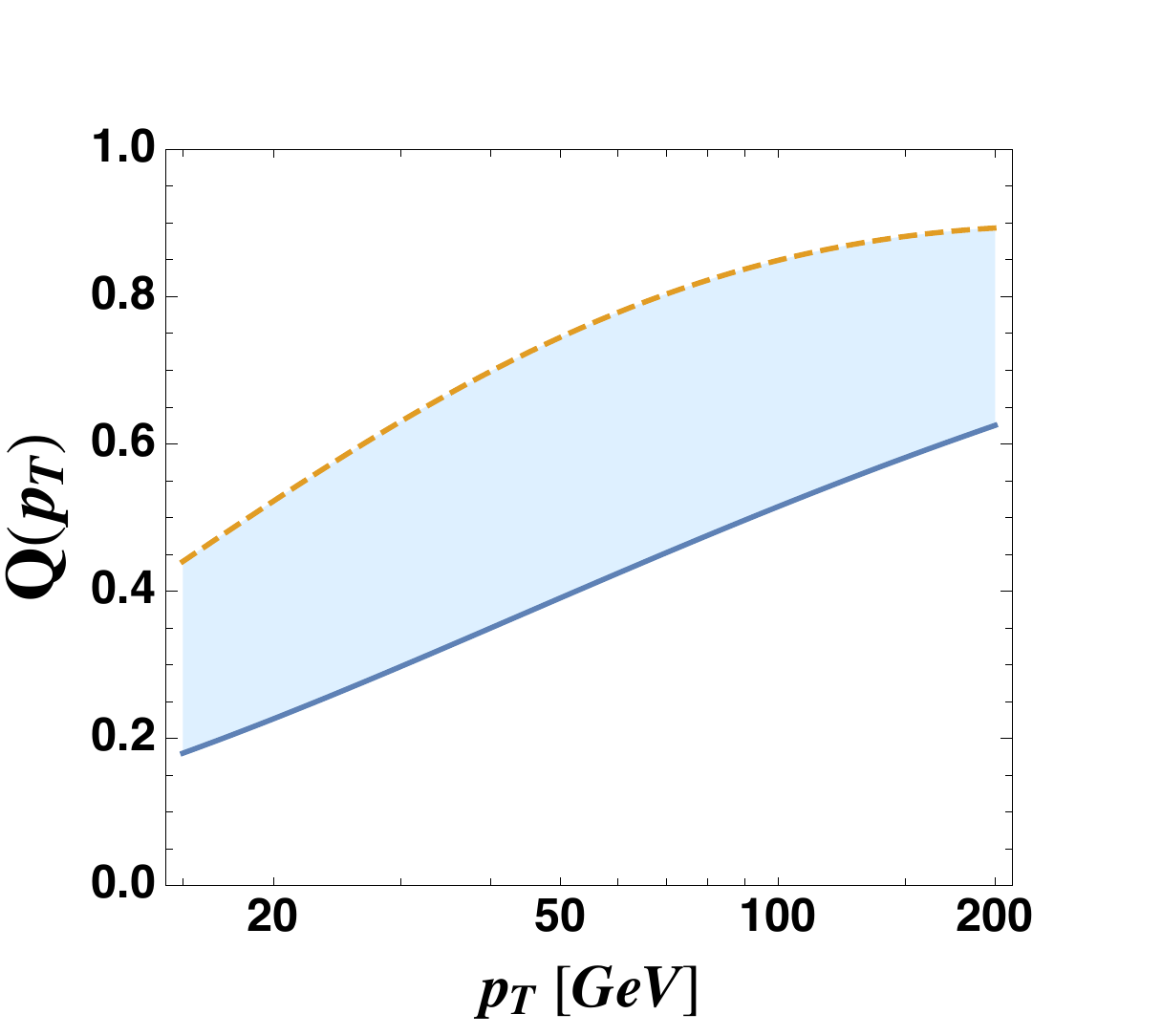}%
\caption{The quenching factor for massless quark $\Qc(\pT|m=0)$ that enters the calculation of the collimator function. The full (lower) line corresponds to the quenching factor with the leading BDMPS-Z soft-gluon spectrum, cf. first term in \eqref{eq:heavy-quark-total-quench}, while the dashed (upper) line contains sub-leading corrections, cf. \eqref{eq:quench-sub-leading}. The shaded area between the curves corresponds to the uncertainty in modeling radiative energy loss. }
\label{fig:numerics1}
\end{figure}

Before we then present our numerical calculations, it is worth emphasizing our approximations in computing the single-quark quenching factors based on the soft-gluon approximation to the full BDMPS-Z spectrum. It was already pointed out in Ref.~\cite{Baier:2001yt}, that the sub-leading logarithmic and numerical factors play an important role for computing the right order of quenching effects and this was also adopted in \cite{Dokshitzer:2001zm}. For the moment we focus only on the effect of massless quark quenching, which enters the dynamics of the collimator functions, cf. \eqref{eq:quenching-fac-higher-order}. Keeping these corrections, the massless quenching factor in \eqn{eq:heavy-quark-total-quench} should read
\beq
\label{eq:quench-subleading}
\Qc(\pT|m=0) = \exp [ -2 \abar (\sqrt{\pi \hat q L^2 n /\pT} - \ln( 2) \ln (\hat q L^2 n/(2\pT)) - 1.84146)] \,.
\eeq
The sub-leading terms result in a faster approach of the quenching factor to unity. We have plotted the quenching factors for massless quarks that enter the calculation of the collimator function in \autoref{fig:numerics1}, where the parameters were chose as described below and the uncertainty arising from modeling radiative energy loss is marked with the shaded region.

As we have done throughout, we will assume the medium to be static and described by averaged parameters $\hat q$ and $L$. We have chosen $ \hat q = 1 \text{ GeV}^2/\text{fm} $ and $L = 2.5 $ fm, and chosen $\abar = 0.15$ \footnote{Note also that the value of the strong-coupling constant $\alpha_s$ can, in principle, be different in the quenching factor and in the collimator function or, more precisely, the running takes place at different scales: in the quenching factor with the typical medium transverse scale and in the collimator with typical jet $k_\perp$.}. These are are qualitatively in the same range as the values obtained in more sophisticated extractions from comparisons to experimental data. Furthermore, we compute the collimator function $R=0.4$ jets and the power of the steeply falling heavy-quark spectrum $n=5$, that was extracted from a fit of the jet data \cite{Cacciari:2018web}. With these choices $m_\ast \simeq 1.5 $ GeV, and therefore the charm quark ($m = 1.3$ GeV) can be considered light while the bottom quark ($m = 5$ GeV) is heavy.

\begin{figure}[t]
\centering
\includegraphics[width=0.45\textwidth]{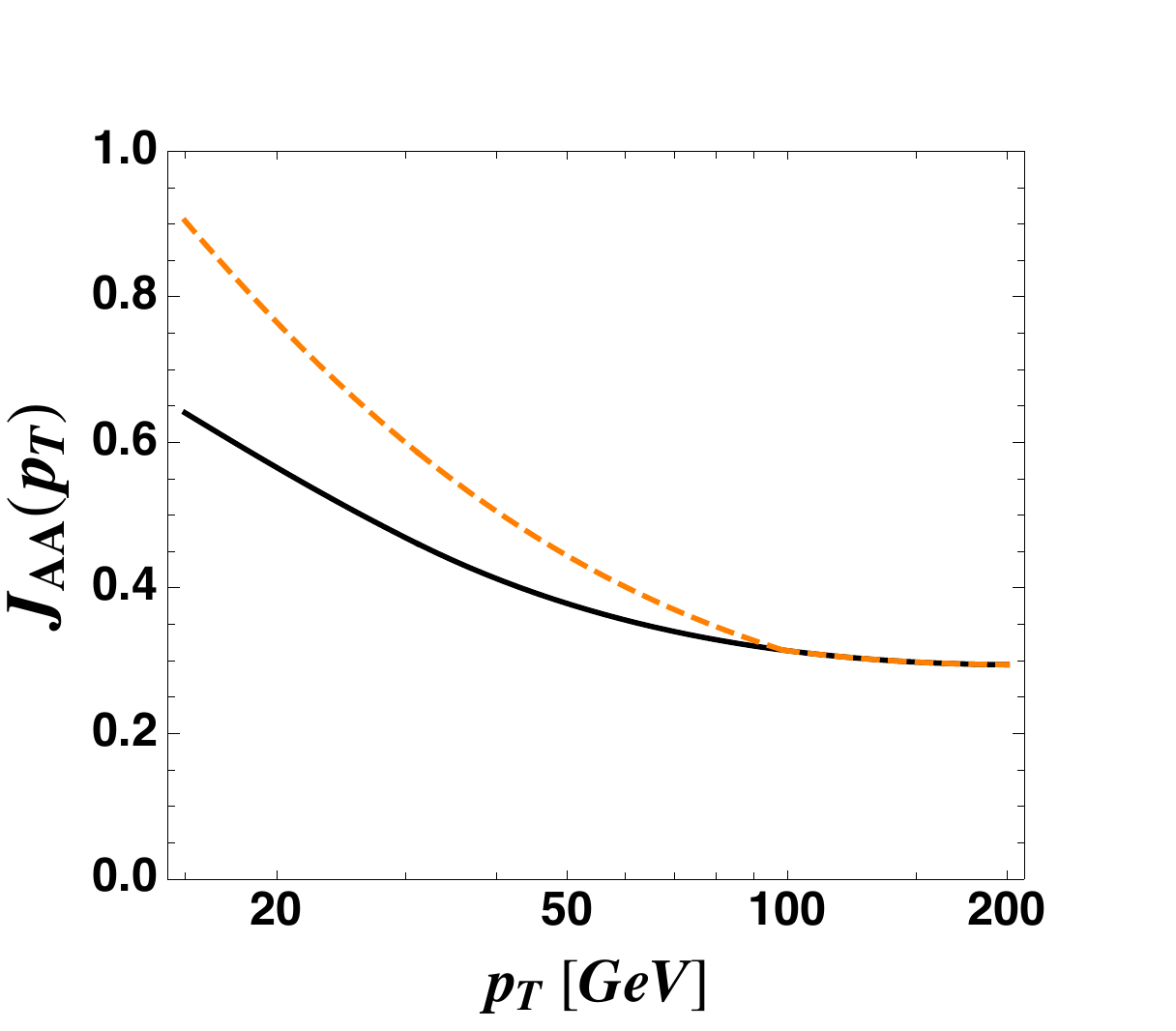}%
\includegraphics[width=0.45\textwidth]{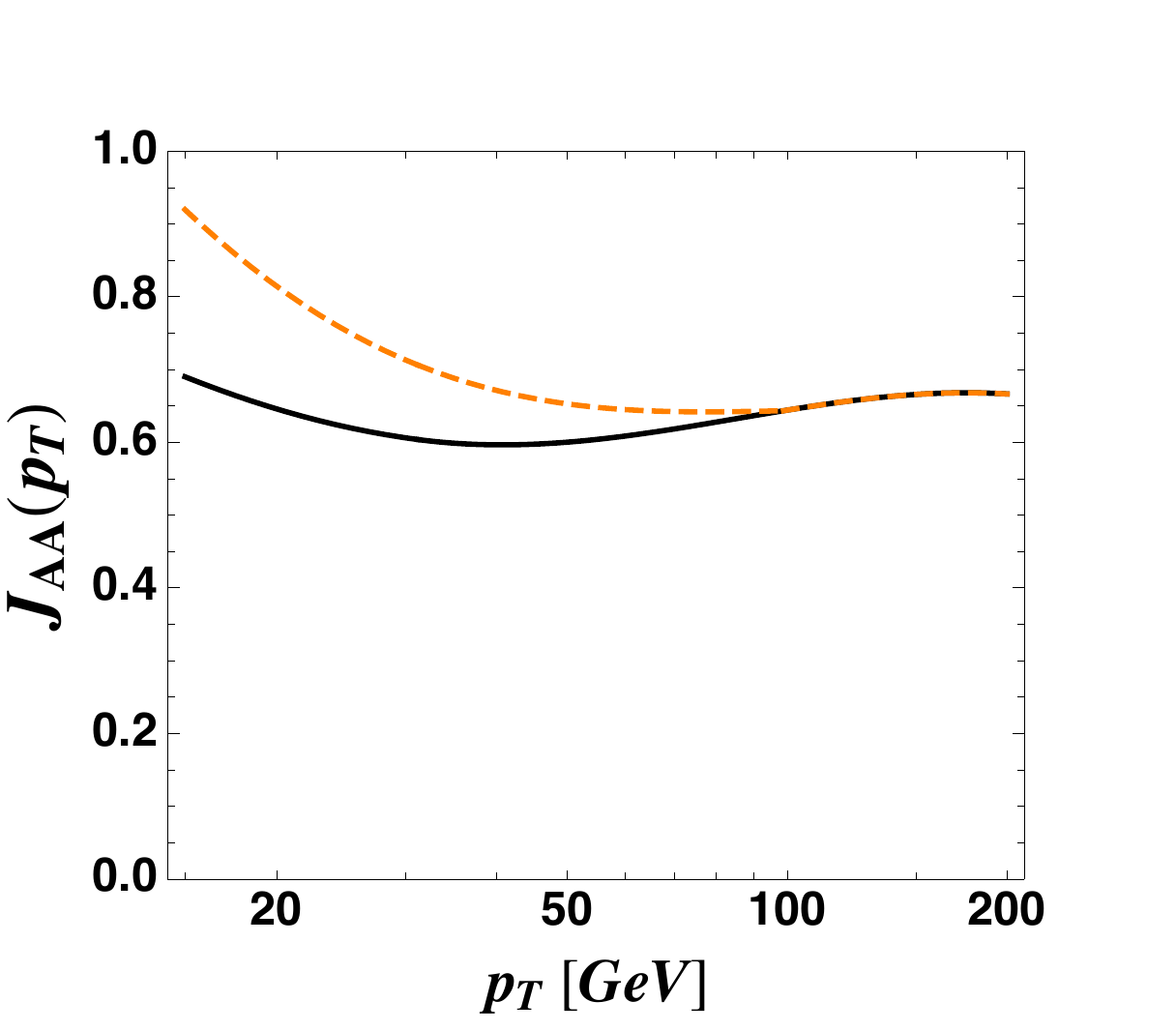}
\caption{The ratio \eqn{eq:osm} as a function of energy for a light quark (charm, $m = 1.3$ GeV) (solid, black line) and heavy quark (bottom, $m=5$  GeV) (dashed, orange line). On the left we have used only the leading term of soft-gluon approximation of the BDMPS-Z spectrum; on the right side, we included numerical corrections to the quenching, as in Refs.~\cite{Baier:2001yt,Dokshitzer:2001zm}.}
\label{fig:numerics}
\end{figure}

We have adopted the approximations and the two ways of estimating the associated, massless quenching factors as discussed above in computing $J_\text{\tiny AA} (\pT) $ in \autoref{fig:numerics}. As a result, the ratio is simply given by the collimator functions that have been explicitly computed to next-to-leading order in Eqs.~\eqref{eq:m1}--\eqref{eq:m3}, with the associated quenching factors computed in \autoref{fig:numerics1}, and whose full resummation is given in \eqn{eq:m4}. The solid, black (dashed, organe) curves represent the collimator function for charm (bottom) quark jets.
On the left side, we used the massless quenching factor associated with the first leading term of the BDMPS-Z soft-gluon spectrum while, on the right side, we included the sub-leading logarithmic and numerical factors, as was done in Refs.~\cite{Baier:2001yt,Dokshitzer:2001zm}. 
It is clear from \autoref{fig:numerics} that the order of magnitude of the effect does depend on the inclusion of these sub-leading BDMPS-Z corrections since the full expression \eqref{eq:quench-subleading} approaches unity much faster than the leading term alone. However, the characteristic $\pT$ where the collimator function of charm and bottom starts deviating on the level of $\sim 10 \%$ is roughly the same in the two cases (for our choice of parameters, this corresponds to roughly $\pT \sim 50$ GeV).

\section{Conclusions}
\label{sec:conclusions}

We have calculated a subset of higher-order corrections for massive-quark jet propagating in the quark-gluon plasma that are enhanced by logarithms of the jet energy. We considered the corrections due to rapid split of the leading particle into hard dipoles well within the medium. This contribution also plays an important role in the context of jet substructure \cite{Casalderrey-Solana:2017mjg}. We have shown that these corrections lead to the enhancement of jet energy loss, and consequently to the decrease of the jet quenching factor, like it was found for the case of massless quark/gluon jets previously. 

The additional suppression can be factorized into a collimator function, that enhances the quenching factor associated with the leading particle and is here evaluated in the leading-logarithmic asymptotics. We have demonstrated that  these corrections are essentially determined by phase space restrictions available for dipole creation --- on the one hand related to the criterium of early splitting, in particular with formation times $\tform < \tdecoh $, and, on the other hand, at angles larger than the associated dead-cone angle. These semi-quantitive estimates show that these corrections are of the same order of magnitude as for massless quark at high-$\pT$, contrary to leading order results \cite{Dokshitzer:2001zm}, where the substantial difference between quenching factors of massless and heavy quark jets was found at relatively low $\pT$. The reason is that in significant part of the parameter space the heavy and massless quark corrections are just the same, and in the remaining region they only differ by the argument of the logarithm.

Our main results are given by Eqs.~\eqref{eq:m2} and \eqref{eq:m3}, where we demonstrate that, from the point of view of higher corrections, we can divide the quarks into \emph{light}, with $m < (qL)^\onehalf$ and \emph{heavy}, $m >  (qL)^\onehalf$. For light quarks the corrections are exactly the same as for the massless quark, see \eqn{eq:m1}, and start to be weakly mass dependent only for rather small energies of order $\pT \leq m^4/\hat q$. For heavy quarks, given in \eqn{eq:m3}, already at large $\pT > \omega_c$ the dead-cone effect start to play role.

 Our work shows that  antenna corrections are essentially the same for massless and heavy quarks, leading to correct from the experimental point of view decrease of quenching factors of heavy quark jets relative to the Dokshitzer-Kharzeev result  \cite{Dokshitzer:2001zm}. Consequently they had little influence on the problem first raised in \cite{Dokshitzer:2001zm}, namely that, contrary to calculation focussing on radiative energy loss, there is not much difference between heavy and massless quark quenching factors experimentally \cite{Sharma:2011zz,Wang:2017vrn} (although the situation can improve in the future \cite{CMS:2017dec}). Nevertheless, since the BDMPS-Z calculation is the leading mechanism available for jet energy loss at high-$\pT$ it is of great interest to test it for different parts of the parameter space.
 
We have also derived the spectra and rates related to a color-charged, massive dipole formed early in the medium, see \eqn{eq:final-spectrum-1a} and \eqref{eq:final-spectrum-1b}, and Eqs.~\eqref{eq:rate-gluon}, \eqref{eq:massive-spectrum-direct-scaling}, \eqref{eq:Int1Final} and \eqref{eq:Int2Final}.

We did not include the restrictions on the phase space that were found to be important for calculation of massless and massive quark quenching weights, see \cite{Salgado:2003gb} and \cite{Armesto:2005iq,Armesto:2003jh}. We do not expect that these restrictions will make qualitative influence on our results. The reason is that the corrections that we calculated were dominated by small frequencies regime outside the dead-cone while the corrections discussed in \cite{Armesto:2005iq,Armesto:2003jh} are mostly important for frequencies in, or close to, the dead-cone region. Besides, the analysis of these constraints needs detailed numerical investigation.

Our results are valid in the leading-logarithmic approximation at high energies. Going to higher logarithmic precision and for applications at very high energies we expect the corrections to satisfy nonlinear integral equations similar to the one that was suggested in  \cite{Mehtar-Tani:2017web}. In this context, the effects of the dead-cone suppression in secondary heavy-quark production, i.e. from the splitting of a gluon into a massive $q\bar q$ pair, are still largely unexplored. For a clearer interpretation, these contributions could perhaps be suppressed using state-of-the-art grooming techniques, employed in \cite{Cunqueiro:2018jbh}.

The results worked out here demonstrate the factorization between the leading term in the quenching weight, calculated in  \cite{Dokshitzer:2001zm} and higher order corrections, and introduces a new way to address jet observables involving massive quark, e.g. the ``leading particle effect'' \cite{Bjorken:1977md,Azimov:1982ef,Azimov:1983ia}, that we plan to address in forthcoming works. From a phenomenological point of view, it will be interesting to study the ratio of quenching factors of heavy quark jet and heavy mesons as a check of the current approach, cf. \eqn{eq:osm}.

\acknowledgements{
KT is supported by a Starting Grant from Bergen Research Foundation (``Thermalizing jets: novel aspects of non-equilibrium processes at colliders'') and the University of Bergen.
BB was supported by the Israel Science Foundation (ISF) grant no. 2025311. The authors thank the CERN Theory Department for hospitality, where most of the work was done.
}

\appendix
\section{Derivation of the interference spectrum}
\label{sec:antenna-spectrum}

Let us derive the rates of gluon emissions of massless and massive particles. Consider a parton splitting at very short time-scales that quasi-instantaneously forms a dipole inside the medium. Using conventional arguments in perturbation theory, one can thereafter define a subsequent emission spectrum off this system. For further applications, it will be sufficient to calculate the interference spectrum between the two constituents of the dipole since direct emissions can be recovered by setting the dipole opening angle to zero.

For concreteness, and in order to clarify the color structure of the process, consider a color-charged dipole originating from a $q (\vec p_0) \to q(\vec p_1) + g(\vec p_2)$ splitting, where we have identified the momentum flow of the splitting \footnote{The notation $\vec p =(E,\p)$ involves the light-cone energy variable $E \equiv (p^0 + p^3)/2$ and transverse momentum $\p = (p^1,p^2)$. Similarly, all time coordinates (and lengths) refer to the light-cone variable $t \equiv (x^0 + x^3)/2$}. We assume, for the time being, that the gluon is massive; this allows us to easily generalize the formula for arbitrary dipoles. The amplitude describing the emission of a soft gluon inside a medium of size $L$ is the sum of two terms that correspond to the possibility of being radiated by dipole constituents. They are given by
\begin{align}
\label{eq:hq-amplitude-1}
\Mc_1 &\sim \frac{g}{\omega} \int_0^L \dd t \, \rme^{i\frac{\omega  }{2 }(\n_1^2 + \Theta_1^2)t}\,\,  (\bdel_x + i \omega \n_1 ) \cdot \beps^\ast_\lambda\, \left. \Gc^{ab} (\k,L; \x, t ) \right|_{\x=\n_1 t} \nn
&\times \left[  V_{\n_1} (L,t)    \tmat^b V_{\n_1}(t, 0)  \right]_{ij} U_{\n_2}^{dc}(L,0) \tmat^c_{jk} \Jc_k(\vec p_0)\,, \\
\label{eq:hq-amplitude-2}
\mathcal{M}_2 &\sim \frac{g}{\omega} \int_0^L \dd t \, \rme^{i\frac{\omega  }{2 }(\n_2^2 + \Theta_2^2)t}\,\,  (\bdel_x + i \omega \n_2 ) \cdot \beps^\ast_\lambda\, \left. \Gc^{ab} (\k,L; \x, t ) \right|_{\x=\n_2 t} \nn
&\times V_{\n_1}(L, 0)_{ij} \left[U_{\n_2}(L,0) \Tmat^b U_{\n_2}(t,0) \right]^{dc} \tmat^c_{jk} \Jc_k(\vec p_0) \,,
\end{align}
up to factors that cancel in the cross section and where the fundamental color matrix $\tmat^c_{jk}$ accounts for the color conservation of the system (for completeness, we recall that $[\Tmat^b]^{ac} \equiv i f^{abc}$). The color factor related to the dipole splitting will be factored in the final expression of the emission spectrum. The dead-cone angles are denoted $\Theta_i \equiv m_i/E_i$ and $\n_i \equiv \p_i/E_i$ determine the trajectories of the dipole constituents. Finally, $\mathcal{J}_{k}(\vec p_0)$ represents the initial quark current.

Medium interactions can be encapsulated into color rotation matrices. To keep explicit track of the color we denote by $V_\n(t,0) = \mathcal{P} \exp [ ig \int_0^t \dd s \, \tmat \cdot  \mathcal{A}(s,\n s)]$ a (path-ordered) Wilson line in the fundamental representation that resums interactions with the medium that is modeled by a background field $\mathcal{A}(t,\x)$. Similarly, $U_{\n}(t,0)$ denotes a Wilson line in the adjoint representation, with the substitution $\tmat^b \to \Tmat^b$. In Eqs. \eqref{eq:hq-amplitude-1} and \eqref{eq:hq-amplitude-2} these objects describe the propagation of the dipole constituents through the medium along fixed trajectories. In contrast, a soft gluon emission in the medium can experience momentum broadening due to transverse momentum exchanges with medium constituents. These interactions are encapsulated in the dressed propagator $\Gc^{ab}$, that is given by
\beq
\Gc^{ab}(\x_1,t_1;\x_0,t_0) = \int_{\r(t_0) = \x_0}^{\r(t_1) = \x_1} \Dc \r \, \exp \left[ i \frac{E}{2} \int_{t_0}^{t_1} \dd t \, \dot \r^2(t) \right] U_{\r(t)}^{ab}(t_1,t_0) \,,
\eeq
where the trajectory of the gluon $\r$ is explicitly time-dependent.

\begin{figure}[t!]
\centering
\includegraphics[width=0.4\textwidth]{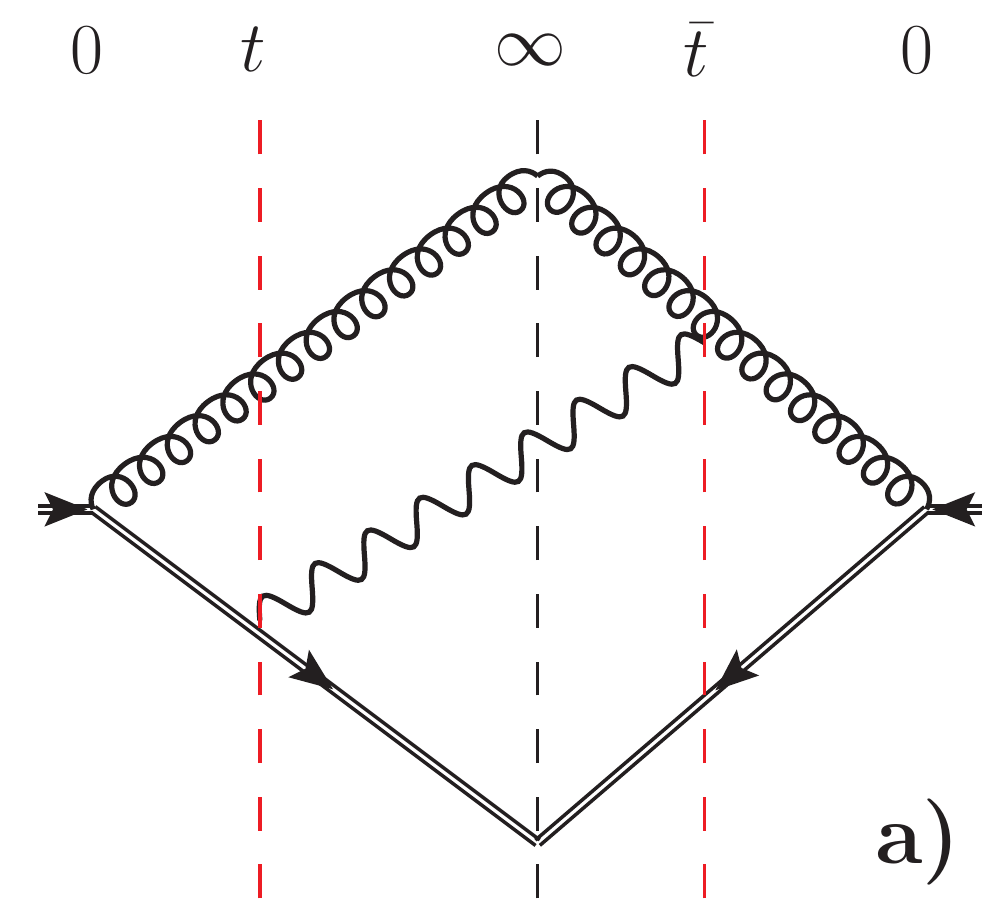}
\quad
\includegraphics[width=0.4\textwidth]{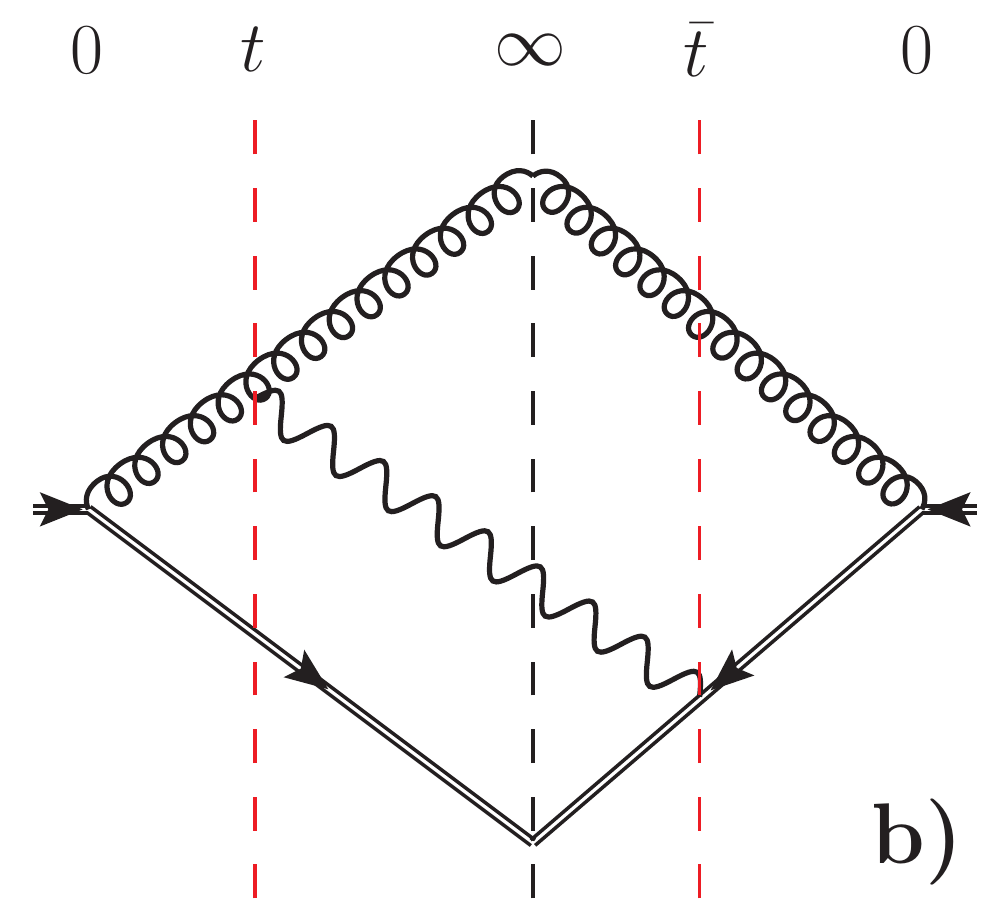}
\caption{Two contributions to the interference spectrum of a heavy-quark--gluon dipole, where the (dressed) propagators of the hard quark (gluon) are represented by a double-line (spiral) while the soft, medium-induced gluon emission is represented by a wavy line. We draw the diagrams for $\bar t > t$.}
\label{fig:diagram}
\end{figure}

We proceed now with the calculation of the interference spectrum. It was first computed for a color-charged dipole involving a massive quark in \cite{Calvo:2014cba} but an explicit expression is not available in the literature. It is therefore meaningful to rederive the spectrum in more generality here. The spectrum involves in total four terms, corresponding to the different time orderings of emission (absorption) in the amplitude and the complex-conjugate amplitude. 
We depict the two main possibilities in \autoref{fig:diagram}. The process in \autoref{fig:diagram}a depicts an emission from the heavy-quark (in the amplitude) and later absorption by a gluon (in the complex conjugate), while the second diagram, \autoref{fig:diagram}b, describes an emission from gluon and subsequent absorption by the heavy-quark. The remaining contributions can be found by adding the complex conjugate of these terms to the final answer, and will be automatically included in the expressions below.

We now have for the diagram \autoref{fig:diagram}a where gluon is emitted at time $\ti$ and absorbed at a later time $\tf$, i.e. $\tf > \ti$, by a heavy quark. Then, the double-differential spectrum reads
\begin{align}
\label{eq:A3}
\left. \frac{\dd I_\text{int}}{\dd \Omega_k} \right\vert_a
&= \frac{1}{N_c C_F} \frac{g^2}{\omega^2} \, 2 \rmR \int^L_0\dd \ti \int ^L_{\ti} \dd\tf \, \rme^{-i\frac{\omega}{2} (\n_2^2+\Theta_2^2) \tf + i\frac{\omega}{2} (\n_1^2 + \Theta_1^2)\ti } (\bdel_{\bar x} - i \omega \n_2) \cdot (\bdel_x + i \omega \n_1) \nn
&\times \rmTr \Sigma \,\int_{\u,\y_1,\y_2} \,\rme^{-i \k \cdot(\y_2 - \y_1)} \frac{1}{N_c^2-1} \langle \rmTr \, \Gc^\dagger (\bar \x, \tf; \y_1,L) \Gc(\y_2,L; \u,\tf)\rangle \nn
&\times \frac{1}{N_c^2-1} \langle \rmTr \, U_{\n_2}^\dagger(\ti,\tf) \Gc(\u,\tf; \x,\ti) \rangle \frac{1}{N_c^2-1} \langle \rmTr \, U_{\n_2}^\dagger(0,\ti) U_{\n_1}(\ti,0) \rangle
\big\vert_{\substack{\x = \n_1 \ti \\ \bar \x = \n_2 \tf}} \,,
\end{align}
where $\dd \Omega_k = \dd \omega \dd^2 \k/[(2\pi)^3 2 \omega]$ is the Lorentz-invariant phase-space element, $\rmTr \Sigma \equiv i f^{a b c} \rmtr (\tmat^a \tmat^b \tmat^c)= -N_c(N_c^2-1)/4$ and $2 \rmR$ accounts for the two possible time orderings of this process, and we have used the short-hand notation $\int_\u \equiv \int \dd^2 \u$.
The normalization factor $(N_c C_F)^{-1}$ takes care of the averaging over the colors of the initial quark and divides out the color structure related to the dipole splitting. The in-medium two-point correlators appearing in \eqref{eq:A3} are known from the literature and, after some simplifications, we rewrite \eqref{eq:A3} as
\begin{align}
\label{eq:A4}
\left. \frac{\dd I_\text{int}}{\dd\Omega_k} \right\vert_a
&= -\frac{N_c}{2} \frac{g^2}{\omega^2} \, 2 \rmR \int^L_0\dd \ti \int ^L_{\ti} \dd\tf \int_\u\, \rme^{-i \k \cdot \u - \frac{N_c n}{2}(L-\tf) \sigma(\u)}\rme^{-i\frac{\omega}{2} (\n_2^2+\Theta_2^2) \tf + i\frac{\omega}{2} (\n_1^2 + \Theta_1^2)\ti } \,\Sc_2(\ti) \nn
&\times(\bdel_{\bar x} - i \omega \n_2) \cdot (\bdel_x + i \omega \n_1)  \, \Kc(\bar \x, \x) \rme^{-i \frac{\omega}{2} \n_2^2(\tf- \ti) + i \omega \n_2 \cdot (\bar \x - \x)} \big\vert_{\x = \n_1 \ti, \bar \x = \u+ \n_2 \tf} \,,
\end{align}
where
\beq
\Sc_2(\ti) = \exp \left[ - \frac{N_c n}{2} \int_{\ti}^{\tf} \dd s\, \sigma(\n_{12}s) \right]
\eeq
and 
\beq
\Kc(\bar \x,\x) = \int_{\r(\ti) = \x}^{\r(\tf) = \bar \x} \Dc \r \, \exp \left[ \frac{i \omega}{2}\int_{\ti}^{\tf} \dd s \, \dot \r^2 - \frac{N_c n}{2} \int_{\ti}^{\tf} \dd s\, \sigma(\r) \right] \,,
\eeq
where we have assumed that the medium is described by a static density $n$.
The overall color factor in \eqref{eq:A4}, $-N_c/2$, corresponds to the correct (negative) interference charge.  
After performing the derivatives and shifting the coordinates, we finally obtain
\begin{align}
\label{eq:A5}
\left. \frac{\dd I_\text{int}}{\dd\Omega_k} \right\vert_a
&= -\frac{N_c}{2} \frac{g^2}{\omega^2} \, 2 \rmR \int^L_0\dd \ti \int ^L_{\ti} \dd\tf \, \rme^{-i \k \cdot \u - \frac{N_c n}{2}(L-\tf) \sigma(\u)} \rme^{i\frac{\omega}{2}\n_{12}^2 \ti} \rme^{ i\frac{\omega}{2} ( \Theta_1^2\ti - \Theta_2^2 \tf) } \,\Sc_2(\ti) \nn
&\times( \bdel_x + i \omega \n_{12}) \cdot \bdel_{\bar x}  \, \Kc(\bar \x, \x) \big\vert_{\bar \x = \u, \x = \n_{12} \ti } \,.
\end{align}
The calculation of the second diagram in \autoref{fig:diagram}b is completely analogous. From symmetry, it turns out that $\dd I / \dd \omega \vert_b = \dd I / \dd \omega \vert_a (\Theta_1 \leftrightarrow \Theta_2)$.  Hence, summing up all four contributions, our final result for the emission spectrum therefore reads,
\begin{align}
\label{eq:final-spectrum-1a}
\frac{\rmd I_\text{int}}{\rmd \Omega_k} &= - \frac{N_c}{2} \frac{g^2}{\omega^2}\,  2\rmR\,  \int_0^{\infty}\dd \ti \int_\ti^{\infty} \dd \tf \int \dd^2 \u \, \rme^{-i \k \cdot \u - \frac{N_c n}{2}(L-\tf) \sigma(\u)} \nn
& \times \rme^{i \frac{\omega}{2} \n_{ij}^2 \ti }\left[\rme^{ i \frac{\omega}{2} (\Theta_i^2 \ti - \Theta_j^2 \tf) }+\rme^{ i \frac{\omega}{2} (\Theta_j^2 \ti -\Theta_i^2 \tf) } \right] \,\Sc_2(t) \nn
&\times  (\bdel_{x} + i\omega \n_{ij})\cdot \bdel_{\bar x} \, \Kc(\bar\x,\x)\big\vert_{\bar \x=\u,\x=\n_{ij} \ti} \,.
\end{align}
We have  checked that this formula reproduces the double-differential interference spectrum in vacuum. It was first discussed in Ref.~\cite{Calvo:2014cba}.  Note, however, that our formula differ from the related Eqs. (4.1) and (4.2) in Ref.~\cite{Calvo:2014cba} by the second term in the square brackets.

The in-medium energy spectrum is found by integrating out the transverse momentum, in which case we obtain
\begin{align}
\label{eq:final-spectrum-1b}
\frac{\rmd I_\text{int}}{\rmd \omega} &= - \frac{N_c}{2} \frac{\alpha_s }{\omega^3}\,  2\rmR\,  \int_0^{\infty}\dd \ti \int_\ti^{\infty} \dd \tf  \, \rme^{i \frac{\omega}{2} \n_{12}^2 \ti }\left[\rme^{ i \frac{\omega}{2} (\Theta_1^2 \ti - \Theta_2^2 \tf) }+\rme^{ i \frac{\omega}{2} (\Theta_2^2 \ti -\Theta_1^2 \tf) } \right]\, \Sc_2(t) \nn
&\times (\bdel_{x} + i\omega \n_{12})\cdot \bdel_{\bar x} \, \Kc(\bar\x,\x)\big\vert_{\bar \x=0,\x=\n_{12} \ti} \,,
\end{align}
which is the main formula for analyzing the direct and interference rates in this work.

\section{Direct and interference radiation rates in the multiple-soft scattering approximation}
\label{sec:direct-interference-rates}

The formulas in Eqs.~\eqref{eq:final-spectrum-1a} and \eqref{eq:final-spectrum-1b} can easily be generalized to any splitting process involving a gluon emission (for the time being let us disregard photon splitting into quark-antiquark). The general formula for the energy spectrum, cf. \eqn{eq:final-spectrum-1b}, reads,
\begin{align}
\label{eq:interference-spectrum-1}
\frac{\rmd I_{ij}}{\rmd \omega} &=\frac{\alpha_s }{\omega^3} Q_{ij}\,  2\rmR\,  \int_0^{\tend}\dd \ti \int_\ti^{\tend} \dd \tf\, 
\rme^{ i \frac{\omega}{2} \n_{ij}^2 \ti } \left[\rme^{ i \frac{\omega}{2} (\Theta_i^2 \ti - \Theta_j^2 \tf) }+\rme^{ i \frac{\omega}{2} (\Theta_j^2 \ti -\Theta_i^2 \tf) } \right]\,\Sc_2(t) \nn
&\times (\bdel_{y} + i\omega \n_{ij})\cdot \bdel_{x} \, \Kc(\x,\y)\big\vert_{\x=\0,\y=\n_{ij} \ti} \,,
\end{align}
where $Q_{ij} \equiv \Q_i \cdot \Q_j = (\Q_0^2 - \Q_1^2 - \Q^2_2)/2$ and $\Q_i$ corresponds to the color charge-vector of the emitter (e.g. for a quark $\Q_q^2 = C_F$ and for a gluon $\Q_g^2 = N_c$). The particle $i$ is associated to the dead-cone angle $\Theta_i$ and direction $\n_i = \p_i/E_i$ (and similarly for particle $j$). The two terms describe two possible emission processes, namely one that is initiated by particle $i$ and one initiated by particle $j$. The case under study, the $Q \to Q+g$ splitting that contributes to the fragmentation of a heavy-quark jet, corresponds to $\Theta_j = 0$ and $Q_{ij} = - N_c/2$, as derived explicitly in \eqref{eq:final-spectrum-1a}.

In \eqref{eq:interference-spectrum-1}, $\Sc_2(t)$ is a two-point function describing the decoherence of the antenna before the splitting occurs and is often referred to as the decoherence parameter \cite{MehtarTani:2010ma,MehtarTani:2011tz,CasalderreySolana:2011rz,MehtarTani:2012cy}. It depends explicitly on the opening angle of the pair, in particular $\Sc_2(t) = 1$ for $\n^2_{12} = 0$. The formula in \eqref{eq:interference-spectrum-1} generalizes the result previously obtained in \cite{Calvo:2014cba}. It is worth pointing out that, due to the Galilean symmetry of the problem, the spectrum does not depend on the directions of the emitters apart from the dipole opening angle $\n_{12}$.

Working in the multiple-soft scattering limit of medium interactions, the $n$-point functions can be found exactly. 
It corresponds to the following approximation on the interaction term, $N_c n \sigma(\r) \approx \hat q \r^2/2$, where $\hat q$ is the jet quenching parameter.
For the decoherence parameter, we immediately obtain $\Sc_2 (t) = \exp[- \hat q \n_{12}^2 t^3/12]$, and for the three-point function describing interactions during the formation of the soft gluon, we find
\beq
\label{eq:three-point-correlator}
\Kc(\x,\y) = \frac{\omega \Omega}{2\pi i \, \sinh \Omega \tau} \, \exp\left\{ \frac{i\omega \Omega}{4 }\left[\tanh \frac{\Omega\tau}{2} (\x+\y)^2+ \coth \frac{\Omega\tau}{2}(\x-\y)^2\right]\right\},
\eeq
where $\tau \equiv \tf - \ti$ and $\Omega \equiv (1+i)\sqrt{\hat q/\omega}/2$. The propagator in \eqref{eq:three-point-correlator} constraints the extent of 
the time difference between gluon emission and absorption $\tau \lesssim \sqrt{\hat q/\omega} \equiv \tbr$. Hence, in the limit $\tbr \ll L$ which holds for soft gluon emissions $\omega \ll \hat q L^2$  that are  most important for energy loss at not too high energies \cite{Baier:2001yt}, we can approximate the time integration over the time-difference of the emissions in the amplitude and in the complex-conjugate as
\beq
\label{eq:time-approximations}
\int_{\ti}^L \rmd \tf = \int_{0}^{L-\ti} \rmd \tau \approx  \int_{0}^{\infty} \rmd \tau \,,
\eeq
see, e.g., \cite{Blaizot:2012fh}.
Formally, this allows to treat multiple radiation as independent with a constant rate.

Focussing again on a $q \to q+g$ splitting, we therefore obtain the formulas for the direct and interference rates. In order to separate the dynamics of the dipole \emph{before} and \emph{during} the emission-time of medium-induced gluon, we will explicitly define the emission rate $\Gamma_{ij}(\omega,t)$ as
\beq
\frac{\dd I_{ij}}{\dd \omega \, \dd t} = \Sc_2(t) \times \tilde \Gamma_{ij}(\omega,t) \,,
\eeq
where the decoherence parameter is responsible for the long-distance color decoherence processes that are happening \emph{prior} to the emission.
Setting the dipole opening angle to zero, $\n_{12}^2 \to 0$, we find the independent heavy-quark and gluon spectra.
\begin{align}
\label{eq:independent-rate-quark}
\tilde \Gamma_q(\omega,t) &= \frac{\alpha_s C_F}{\omega^3}\,  2\rmR \int_{0}^{\infty}\rmd \tau \, \rme^{-i\frac{\omega}{2} \Tdc^2\tau}  \bdel_{x}\cdot\bdel_{y}\, {\cal K}(\x,\y)\Big|_{\x=\y=\0} \,, \\
\label{eq:independent-rate-gluon}
\tilde \Gamma_g(\omega,t) &= \frac{\alpha_s N_c}{\omega^3}\,  2\rmR \int_{0}^{\infty}\rmd \tau \, \bdel_{x}\cdot\bdel_{y}\, {\cal K}(\x,\y)\Big|_{\x=\y=\0} \,,
\end{align}
where we have restored the correct mass and color factors, and introduced the definition of the dead-cone angle $\Tdc \equiv \theta_q$ used throughout.
The interference spectrum contains two terms, given by
\begin{align}
\label{eq:interference-rate-term1}
\tilde\Gamma_{12}(\omega,t) &= - \frac{\alpha_s N_c}{2 \omega^3} 2 \rmR \,\rme^{i \frac{\omega}{2} (\n_{12}^2+\Tdc^2)t}\int_0^\infty \dd \tau  \, (\bdel_{y}+i\omega \n_{12}) \cdot \bdel_{x}\, \mathcal{K}(\x,\y) \big|_{\x=0, \y=\x_{12}} \,, \\
\label{eq:interference-rate-term2}
\tilde\Gamma_{21}(\omega,t) &= - \frac{\alpha_s N_c}{2 \omega^3} 2 \rmR \,\rme^{i \frac{\omega}{2} (\n_{12}^2-\Tdc^2 )t}\int_0^\infty \dd \tau  \, \rme^{ -i \frac{\omega}{2}\Tdc^2 \tau}(\bdel_{y}+i\omega \n_{12}) \cdot \bdel_{x}\,\mathcal{K}(\x,\y)  \big|_{\x=0, \y=\x_{12}} \,.
\end{align} 
Furthermore, we can altogether neglect the phases involving the terms $\Tdc^2 t$ since, in the double-logarithmic approximation the hard gluon splitting takes place at angles much larger than the dead-cone angle.

Note that this expression, at finite $\n_{12} \neq 0$, diverges both in the limit of small $t$ and $\tau$. However, this contribution is completely independent of medium parameters, $\sim \omega^2/[ \pi t \tau]$, and can therefore be regularized by subtracting the vacuum contribution. This is in contrast to the regularization of the direct terms, which only exhibit a divergence at small $\tau$, $\sim \omega^2/[\pi \tau]$, but is cured in the same way. It follows that all medium-induced contributions vanish in the $\hat q \to 0$ limit. Hence the rates that are used to compute energy loss processes are therefore regularized as
\beq
\label{eq:regulated-rate}
\Gamma_{ij}(\omega,t) \equiv \tilde\Gamma_{ij}(\omega,t) - \tilde\Gamma_{ij}(\omega,t) \big\vert_{\hat q \to 0} \,.
\eeq
After applying this regularization, we immediately find the rate of direct emissions by a massless gluon,
\beq
\Gamma_g(\omega,t) = \frac{\alpha_s N_c}{\pi} \sqrt{\frac{\hat q}{\omega^3}} \,,
\label{eq:rate-gluon}
\eeq
recovering the well-known LPM rate valid for soft-gluon emission.
For the corresponding rate off a massive quark, we find
\beq
\label{eq:massive-spectrum-direct-scaling}
\Gamma_q(\omega,t) = \frac{\alpha_s C_F}{\pi} \Theta^2 \left[ - \rmI \,\psi_0\left(-\frac{1+i}{4}\sqrt{\frac{\Theta^4 \omega^3}{\hat q}} \right) - \sqrt{\frac{\hat q}{\Theta^4 \omega^3}} - \frac{3}{4}\pi \right] \,,
\eeq
where $\psi_0(x)$ is the digamma function.

Let us continue with the interference spectra. Considering first \eqref{eq:interference-rate-term1}, we define $\alpha' =  \omega t \left(\Tdc^2 + \n_{12}^2 \right)/2 \approx \omega t \n_{12}^2 /2$ and $\alpha  = \alpha' + \omega \x_{12}^2 / (4 \tform)$, where $\tform \equiv \sqrt{\omega /\hat q}$, the first interference term reads simply
\beq
\label{eq:Int1Final}
\Gamma_{12}(\omega,t) = - \frac{\alpha_s N_c}{2\pi \omega \,t } \rme^{-\frac{\omega}{4 \tform} \x_{12}^2} \left[ 2 \rme^{\frac{\omega}{4 \tform} \x_{12}^2} \cos \alpha' -2 \cos \alpha + \frac{t}{\tform} (\cos \alpha - \sin \alpha)\right] \,.
\eeq
Here, putting $\Tdc \to 0$ in the $\alpha$ and $\alpha'$ factors, we recover immediately interference spectrum for a massless antenna.

The second interference term in \eqn{eq:interference-rate-term2} is complicated due to the additional phase factor related to the finite quark mass. 
We define the following integral,
\begin{align}
I_1 &\equiv \int_0^\infty \frac{\dd \tau}{\sinh^2\Omega \tau}\, \rme^{a \coth \Omega \tau + b \tau} =\frac{2}{\Omega} \rme^a \Gamma\left(1-\frac{b}{2 \Omega} \right) U\left(1-\frac{b}{2\Omega},2,-2a \right) \,,
\end{align}
which will become useful later. We have made sure that $\rmR \,b/(2\Omega) >-1$, which allowed us to deform the integration contour to lie along the real axis. We will also need the following integral
\begin{align}
\int_0^\infty \frac{\dd \tau}{\sinh^2\Omega \tau}\, \coth\Omega\tau \,\rme^{a \coth \Omega \tau + b \tau} = \frac{\partial}{\partial a} I_1 = I_1 + I_2 \,,
\end{align}
where 
\beq
I_2 \equiv \frac{4}{\Omega} \rme^a \left(1-\frac{b}{2\Omega}\right)\Gamma\left(1-\frac{b}{2 \Omega} \right) U\left(2-\frac{b}{2\Omega},3,-2a \right) \,.
\eeq
The second interference term can then be written as
\begin{align}
\label{eq:Int2Final}
\Gamma_{21}(\omega,t) &= \frac{\alpha_s N_c}{2 \omega^3} 2 \rmR \,\rme^{i \frac{\omega}{2} (\n_{12}^2 - \Tdc^2 )t} \frac{(\omega\Omega)^2}{2\pi} \left[\left(2 - i \n_{12}^2 t \omega \right) I_1 + i \n_{12}^2 t^2 \omega \Omega \left(I_1 + I_2 \right) \right] \nn
& -\tilde \Gamma_{21}(\omega,t)\big\vert_{\hat q \to 0} \,,
\end{align}
with parameters $a = i  \omega \Omega \n_{12}^2 t^2/2$ and $b = - i \omega \theta_q^2 /2$, 
and where the vacuum subtraction term is simply
\beq
\tilde \Gamma_{21}(\omega,t)\big\vert_{\hat q \to 0} =  \frac{\alpha_s N_c}{2 \omega^3} 2 \rmR \,\rme^{i \frac{\omega}{2} ( \n_{12}^2 - \Tdc^2)t} \frac{\omega^2}{\pi t} \,.
\eeq
We have explicitly checked that taking the massless limit, $\Tdc \to 0$, inside of the square brackets of \eqn{eq:Int2Final} reproduces the first term of the interference  contribution given by  \eqn{eq:Int1Final}.

Although the formulas derived above can be evaluated numerically, we aim at understanding the problem through the fundamental scales appearing in these expressions. As a final step, let us therefore spend some time on discussing the different time-scales appearing, starting with the interferences. Neglecting the dead-cone angle compared to the dipole opening angle, we find that one of the exponentials in Eqs.~\eqref{eq:Int1Final} and \eqref{eq:Int2Final} start oscillating at times
\beq
t  \sim t_1= \frac{1}{\n_{12}^2 \omega} \,.
\eeq
This scale is related to quantum coherence \cite{CasalderreySolana:2011rz,MehtarTani:2012cy}, imposing that the wavelength of emitted quanta resolve the dipole $x_\perp(t) > \lambda_\perp(t) $, where $\lambda_\perp^2 \sim \omega/t$. In the vacuum $t \sim \tform$, which leads immediately to the angular ordering condition. The second time-scale appearing in the exponentials is 
\beq
t_2 = (\hat q \n_{12}^4 \omega)^{-1/4} \,.
\eeq
This scale is not an independent scale since we can rewrite it as $t_2 = (\tdecoh^3 t_1)^{1/4}$, where $\tdecoh$ is the color decoherence scale introduced above. This only allows for two possible orderings, $\tdecoh < t_2 < t_1$ and $t_1 < t_2 < \tdecoh$, and in neither of the cases does $t_2$ constitute the shortest, and therefore most relevant, time-scale.

It is also possible to show that $t_1/t_2 \sim \theta_\text{br}(\omega)/\theta_{12}$, where $\theta_{12} \equiv |\n_{12}|$ and $\theta_\text{br}(\omega) = (\hat q /\omega^3)^{1/4}$ is the typical emission angle for medium-induced gluons.
Hence, since we mainly are interested in large-angle gluon emissions that contribute to energy-loss, $\theta_\text{br}(\omega)> R > \theta_{12}$, $t_1 > t_2$ and the first ordering is actually realized. This means that the decoherence time, which resides in the function $\Sc_2(t)$ sets the shortest time-scale where the interferences will be suppressed.
For our purposes, it is therefore possible to show that we can altogether neglect the phases involving these time-scales in the interference terms, leaving us with
\begin{align}
\label{eq:interference-approximations-1}
\Gamma_{12}(\omega,t) \simeq - \Gamma_g(\omega,t) \Theta\big(\omega - \omega_{12} \big)\,, \\
\label{eq:interference-approximations-2}
\Gamma_{21}(\omega,t) \simeq - \Gamma_q(\omega,t) \Theta\big(\omega - \omega_{12} \big) \,,
\end{align}
where $\omega_{12} \equiv (\hat q/\theta_{12}^4)^{1/3}$.
In Laplace space this becomes,
\begin{align}
\label{eq:interference-laplace-1}
\gamma_{12}(\nu,t) &\simeq - 2\abar \sqrt{\frac{\hat q }{\omega_{12}}} \left[1- \rme^{-\nu \omega_{12}} - \sqrt{\pi \nu \omega_{12}}\, \text{erf} (\sqrt{\nu \omega_{12}})\right] \,, \\
\label{eq:interference-laplace-2}
\gamma_{21}(\omega,t) &\simeq - 2\abar \sqrt{\frac{\hat q }{\omega_\text{min}}} \left[1- \rme^{-\nu \omega_\text{min}} - \sqrt{\pi \nu \omega_\text{min}} \,\text{erf} (\sqrt{\nu \omega_\text{min}})\right] \,,
\end{align}
where $\omega_\text{min} \equiv \min(\oDC,\omega_{12})$ which is simply $\omega_\text{min} = \omega_{12}$ in the leading-logarithmic approximation ($\theta_{12} > \Tdc$). Hence, $\gamma_{12} \simeq \gamma_{21} \approx - 2\abar \sqrt{\hat q} (\sqrt{\pi \nu} - \sqrt{\omega_{12}}) $.

\bibliographystyle{apsrev4-1}
\bibliography{heavyjetquenching}

\end{document}